\documentclass[%
reprint,
superscriptaddress,
preprintnumbers,
bibnotes,
 amsmath,amssymb,
 prd,
]{revtex4-2}

\usepackage{graphicx}
\usepackage{dcolumn}
\usepackage{placeins} 
\usepackage{bm}
\usepackage{booktabs}
\usepackage{xcolor}

\usepackage[colorlinks  = true,
            linkcolor   = blue,
            urlcolor    = blue,
            citecolor   = blue,
            anchorcolor = blue]{hyperref} 
            
\AtBeginDocument{
\heavyrulewidth=.08em
\lightrulewidth=.05em
\cmidrulewidth=.03em
\belowrulesep=.65ex
\belowbottomsep=0pt
\aboverulesep=.4ex
\abovetopsep=0pt
\cmidrulesep=\doublerulesep
\cmidrulekern=.5em
\defaultaddspace=.5em
}

\usepackage{lineno}
            
\begin{document}

\title{Design and performance of a multi-terahertz Fourier transform spectrometer\\ for axion dark matter experiments}

\author{Kristin Dona}\email{kdona@uchicago.edu}\affiliation{Department of Physics, University of Chicago, Chicago IL 60637, USA}
\author{Jesse Liu}\email{jesseliu@uchicago.edu}\affiliation{Department of Physics, University of Chicago, Chicago IL 60637, USA}
\author{Noah Kurinsky}\email{kurinsky@fnal.gov}\affiliation{Kavli Institute for Cosmological Physics, University of Chicago, Chicago IL 60637, USA}\affiliation{Fermi National Accelerator Laboratory, Batavia, Illinois 60510, USA}
\author{David Miller}\email{davemilr@uchicago.edu}\affiliation{Department of Physics, University of Chicago, Chicago IL 60637, USA}

\author{Pete Barry}
\affiliation{Kavli Institute for Cosmological Physics, University of Chicago, Chicago IL 60637, USA}
\affiliation{Argonne National Laboratory, Lemont, IL 60439, USA}
\author{Clarence Chang}
\affiliation{Kavli Institute for Cosmological Physics, University of Chicago, Chicago IL 60637, USA}
\affiliation{Argonne National Laboratory, Lemont, IL 60439, USA}
\author{Andrew Sonnenschein}\email{sonnensn@fnal.gov}\affiliation{Fermi National Accelerator Laboratory, Batavia, Illinois 60510, USA}

\date{\today}


\begin{abstract}
    Dedicated spectrometers for terahertz radiation with [0.3, 30]~THz frequencies using traditional optomechanical interferometry are substantially less common than their infrared and microwave counterparts. 
    This paper presents the design and initial performance measurements of a tabletop Fourier transform spectrometer (FTS) for multi-terahertz radiation using infrared optics in a Michelson arrangement. 
    This is coupled to a broadband pyroelectric photodetector designed for [0.1, 30]~THz frequencies. 
    We measure spectra of narrowband and broadband input radiation to characterize the performance of this instrument above 10~THz, where signal-to-noise is high. 
    This paves the groundwork for planned upgrades to extend below 10~THz. 
    We also briefly discuss potential astroparticle physics applications of such FTS instruments to broadband axion dark matter searches, whose signature comprises low-rate monochromatic photons with unknown frequency. 
\end{abstract}

\maketitle

\section{Introduction}

The Fourier transform spectrometer (FTS) using optomechanical interferometry techniques is often used for spectral analysis of electromagnetic radiation from microwave to visible frequencies. 
The infrared ($f \sim [30, 300]$~THz frequencies) FTS, or FTIR, is a cornerstone of chemistry laboratories~\cite{griffiths2007fourier,NewportFTIR}, with diverse applications from biochemical spectroscopy~\cite{Zanyar2008,geibel2010new,bellisola2012infrared,baker2014using} to atmospheric physics~\cite{gisi2012xco2,cossel2017gas}.
Microwave ($f \sim [30, 300]$~GHz) FTS instruments are widely used for space-based~\cite{Mather1993,naess2019time} and ground-based~\cite{carlstrom201110,Pan:2019omw,thornton2016atacama,matsuda2019polarbear,BICEP2014} cosmology observatories. 
However, dedicated FTS devices for terahertz regimes ($f \sim [0.3, 30]$~THz) are far less common due to the relative immaturity of bright terahertz sources and detectors~\cite{tonouchi2007cutting,zouaghi2013broadband,dhillon20172017,Carelli2017,Martini:2020owl}. 
Nonetheless, terahertz radiation attracts significant community interest due to applications in medicine~\cite{pickwell2006biomedical,reid2010accuracy,shen2011terahertz,Taylor2011}, biomolecular spectroscopy~\cite{Hintzsche2012,williams2013intermolecular,PhysRevLett.106.158102}, atmospheric science~\cite{hindle2008continuous,slocum2013atmospheric,hsieh2016dynamic}, security imaging~\cite{chan2007imaging,Kemp2011,Cooper2014,heinz2015passive}, and telecommunications~\cite{federici2010review,akyildiz2014terahertz,Seeds2015}. 

In astroparticle physics, a longstanding open question is the identity of dark matter (DM), and its laboratory detection using (multi-)terahertz photon signatures is largely unexplored. 
Specifically, bosonic DM candidates such as axions can modify Maxwell's equations to induce monochromatic photons with frequency proportional to the DM mass~\cite{Jaeckel:2010ni,Essig:2013lka,Baker:2013zta,Battaglieri:2017aum,Irastorza:2018dyq}.
Conventional laboratory searches employ resonant microwave cavity techniques~\cite{Sikivie:1983ip,DePanfilis:1987dk,Wuensch:1989sa,Hagmann:1990tj}, where
state-of-the-art experiments such as ADMX~\cite{Asztalos:2001tf,Asztalos:2001jk,Asztalos:2009yp,Wagner:2010mi,Du:2018uak} and HAYSTAC~\cite{Kenany:2016tta,Brubaker:2016ktl,Zhong:2018rsr,Backes:2020ajv} probe axion masses of $[1.8, 24]~\mu$eV  ([0.4, 5.8]~GHz). 
However, this narrowband search technique is impractical for higher masses (frequencies). 
A broadband search strategy is thus desired, where dish antenna techniques show promise~\cite{Horns:2012jf}, but implementing photon detection and spectral analysis across multi-terahertz frequencies poses significant experimental challenges. 
Moreover, the signal rates are suppressed by the small DM--photon coupling, which often require low-noise quantum sensing of single photons~\cite{Ahmed:2018oog,kutas2020terahertz}. 
Constructing a dedicated multi-terahertz FTS is thus motivated to provide a broadband spectral analyzer as groundwork for our planned DM applications~\cite{THzAxionSnowmassLoI,THzAxionSonnenschein}.
This includes instrument calibration, optics characterization, and DM mass reconstruction.

This paper presents the design and initial performance results of a tabletop FTS for analyzing multi-terahertz radiation using off-the-shelf components to keep initial costs low.
We construct a Michelson interferometer utilizing optics manufactured for the mid-infrared and couple this to a photodetector designed for [0.1, 30]~THz frequencies. 
Section~\ref{sec:setup} introduces the experimental setup and alignment along with operational experiences and data-taking methodology. 
Section~\ref{sec:results} presents measurements of narrowband and broadband input radiation together with filter transmission spectra. 
This characterizes the performance of our instrument, where resolution of narrow spectral features is demonstrated for $f \gtrsim 10$~THz with our current setup. 
We discuss future extensions to probe lower frequencies $f \lesssim 10$~THz, testing optics designed for such frequencies along with a more diverse range of sources and detectors. 
Section~\ref{sec:summary} summarizes these results before briefly discussing potential applications of these FTS techniques in astroparticle physics for laboratory DM searches. 

\section{\label{sec:setup}Instrument setup and operation}

\begin{figure}
\centering
\includegraphics[width=\linewidth]{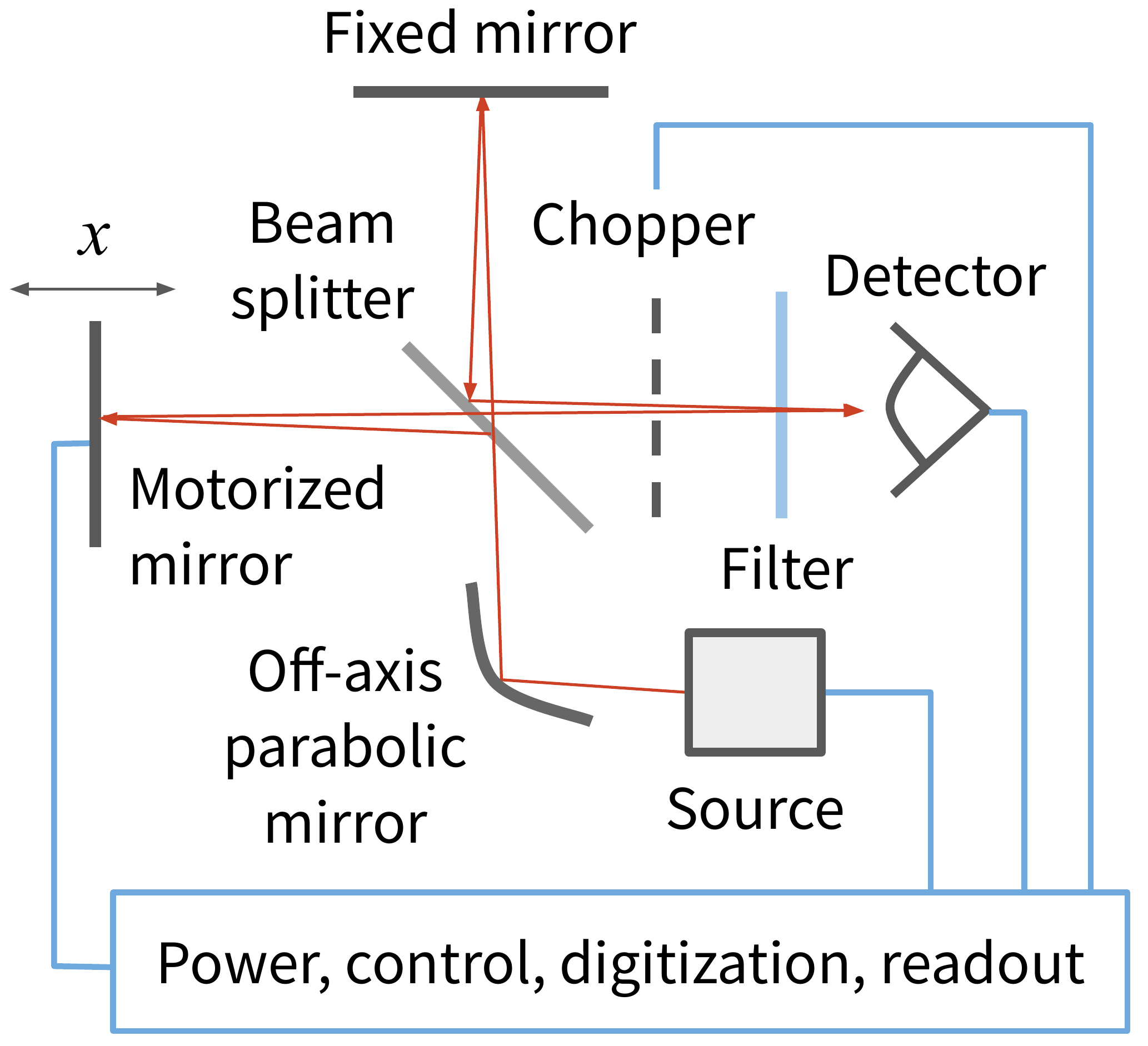}\\[0.3cm]
\includegraphics[width=\linewidth]{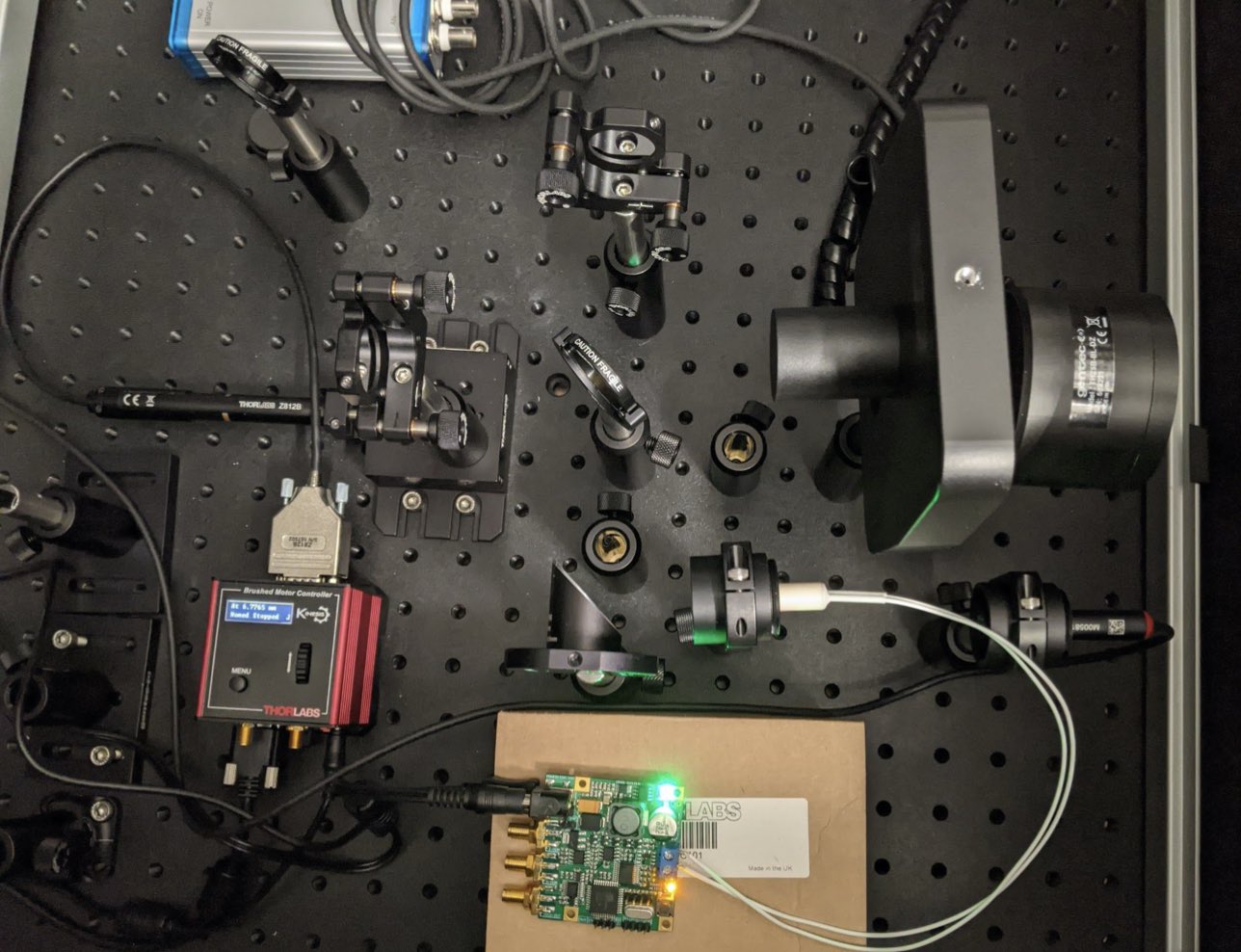}
\caption{\label{fig:michelson}Schematic diagram (top) and corresponding photograph (bottom) of the Fourier transform spectrometer in a Michelson interferometer arrangement. The path of the input radiation is illustrated by the thin red arrows. The optomechanical and backend components are summarized in Table~\ref{tab:interferometer_components} with further details in the main text. }
\end{figure}

The FTS is a Michelson interferometer illustrated in Fig.~\ref{fig:michelson} with subsection~\ref{sec:exp_setup} presenting the components in its assembly.  Subsection~\ref{sec:operation_analysis} then describes the data taking and calculation of the power spectral density.

\subsection{\label{sec:exp_setup}Experimental Setup}

\begin{table}[]
\begin{tabular}{ll}
\toprule
FTS component                 & Attribute                        \\ 
\midrule
Beamsplitter                  & Thorlabs Pellicle BP145B3        \\
\quad R:T datasheet           & [0.4, 2.5] $\mu$m                    \\
\quad Coated for 45:55 R:T    & [1, 2] $\mu$m                          \\
Mirrors                       & Aluminium PF10-03-G01  \\
\quad Design wavelengths      & [0.45, 20] $\mu$m                \\
Fixed arm length              & 76 mm                              \\
Motorized stage               & \\
\quad Model                   & Thorlabs MT1-Z8                  \\
\quad Min.\ step size         & 0.05 $\mu$m                            \\
\quad Max.\ travel distance   & 12 mm                               \\
\midrule
Infrared source               & \\
\quad Model                   & IR-Si253 \\
\quad Emitter material        & Silicon Nitride  \\
\quad Temperature at 9V       & 1200 K\\
\midrule
Photosensor                   & \\
\quad Model                   & Gentec THZ5B-BL-DA-D0 \\
\quad Technology              & Pyroelectric                     \\
\quad Sample rate             & 5 Hz                             \\
\quad Design noise power      & 50 nW                             \\
\quad Design range            & [0.1, 30] THz                     \\ 
\quad Chopper rate            & 25 Hz (model SDC-500)        \\
\quad Readout                 & T-RAD USB 12 bit ADC\\
\midrule
Gentec filter windows         & \\
\quad Polyethylene (PEW)      & [3, 30]~$\mu$m   \\
\quad Silicon (SiW)           & [1.1, 9], [50, 1000]~$\mu$m\\
\bottomrule
\end{tabular}
\caption{\label{tab:interferometer_components}
Summary of optical components, manufacturer and model, together with their design performance specifications used to construct the interferometer, which are illustrated in Fig.~\ref{fig:michelson}. 
Further details about these components are discussed in the main text. }
\end{table}

The components of the FTS are arranged on an optical breadboard (Thorlabs MB2424), which sits on four rubber dampers (Thorlabs RDF1) and is enclosed inside a light-tight box (Newport LTE-22) for environmental noise suppression. 
Table~\ref{tab:interferometer_components} summarizes the components used to construct the interferometer for the studies in this paper, which we discuss in what follows. 

Two classes of electromagnetic radiation sources are considered. First, a narrowband source comprising a USB-powered laser with visible wavelength of 635~nm (Thorlabs PL202) is used to aid and validate optical alignment. 
Second, broadband infrared (IR) radiation is provided by compact thermal radiators manufactured by HawkEye Technologies and supplied by Boston Electronics. 
Our nominal source is a silicon nitride emitter (IR-Si253), which has a design operating temperature of 1420~K at 1.5~A, 12~V, 18~W. 
We also obtained a silicon carbide source (IR-Si207) and a coil-wound source on an aluminium substrate (IR-12K), but their detailed study is deferred to future work.  
For this paper, we power the IR source using a benchtop switching power supply (Circuit Specialists CSI3003SM). 
The IR source is mounted at the focus of a parabolic reflector package, which illuminates a protected-silver-coated off-axis parabolic mirror (Thorlabs MPD129-P01), which directs and collimates the light to the interferometer optics.

The light then reaches a pellicle beamsplitter made from nitrocellulose membrane (Thorlabs BP145B3) designed to minimize multiple reflections. 
This is coated for a 45:55 (reflectance:transmission) split ratio in a design range of [1, 2]~$\mu$m, but we find this also has broad transmission and reflection beyond this spectral range.  
The split light then reaches either a movable or fixed mirror made from protected aluminium (Thorlabs PF10-03-G01), which are situated at the end of the interferometer arms that are approximately 7.6~cm long. 
The mirrors are designed for operation in [0.45, 20]~$\mu$m wavelengths and are placed in kinematic mounts (Thorlabs KM100) to aid precision alignment. 
The movable mirror is mounted on a motorized stage (Thorlabs MT1-Z8), which is specified to have a minimum increment of $\Delta x = 50$~nm and a maximum travel range of 12~mm. 
This motorized stage is connected to the Thorlabs K-Cube motor controller interfaced with the LabVIEW-based \textsc{Kinesis} software package~\cite{ThorlabsKinesis}.

For the detection of broadband photon radiation, we employ the Gentec Electro-Optics pyroelectric radiometer (THZ5B-BL-DA-D0), which operates at room temperature and the manufacturer-specified design frequency range is [0.1, 30]~THz. 
This comprises a 5~mm diameter pyroelectric photosensor designed for integrated relative power measurements down to 50~nW with 1~nW resolution.
The pyroelectric sensor has a dielectric constant with a strong functional dependence on temperature, such that the voltage across the sensor is highly sensitive to absorbed power across a wide range of frequencies. 
In practice, we find the usable frequency range extends above 30~THz, which we demonstrate in section~\ref{sec:results} using different filters and the 635~nm (472~THz) visible laser. 
The manufacturer-specified absorption efficiency~\cite{Gentec-THZ-BL-curves} of the photodetector is relatively high and constant $\sim 90\%$ for $\gtrsim 10$~THz but falls significantly for $\lesssim 10$~THz as shown in the appendix (Fig.~\ref{fig:GentecAbsorption}).
The sensor is coupled to the Gentec T-RAD USB module for readout, which is specified with a 5~Hz sampling rate and 12-bit analog-to-digital converter (ADC) corresponding to a dynamic range of $2^{12} -1$. 
For ambient noise suppression, lock-in amplifier software is used in conjunction with a Gentec SDC-500 optical chopper operating at 25~Hz placed between the beamsplitter and detector. 
To prevent aliasing due to insufficient sampling, the movable mirror must advance sufficiently slowly compared to the 5~Hz detector sampling rate according to Eq.~\ref{eq:sample_frequency} discussed below. 

\begin{figure}
\centering
\includegraphics[width=\linewidth]{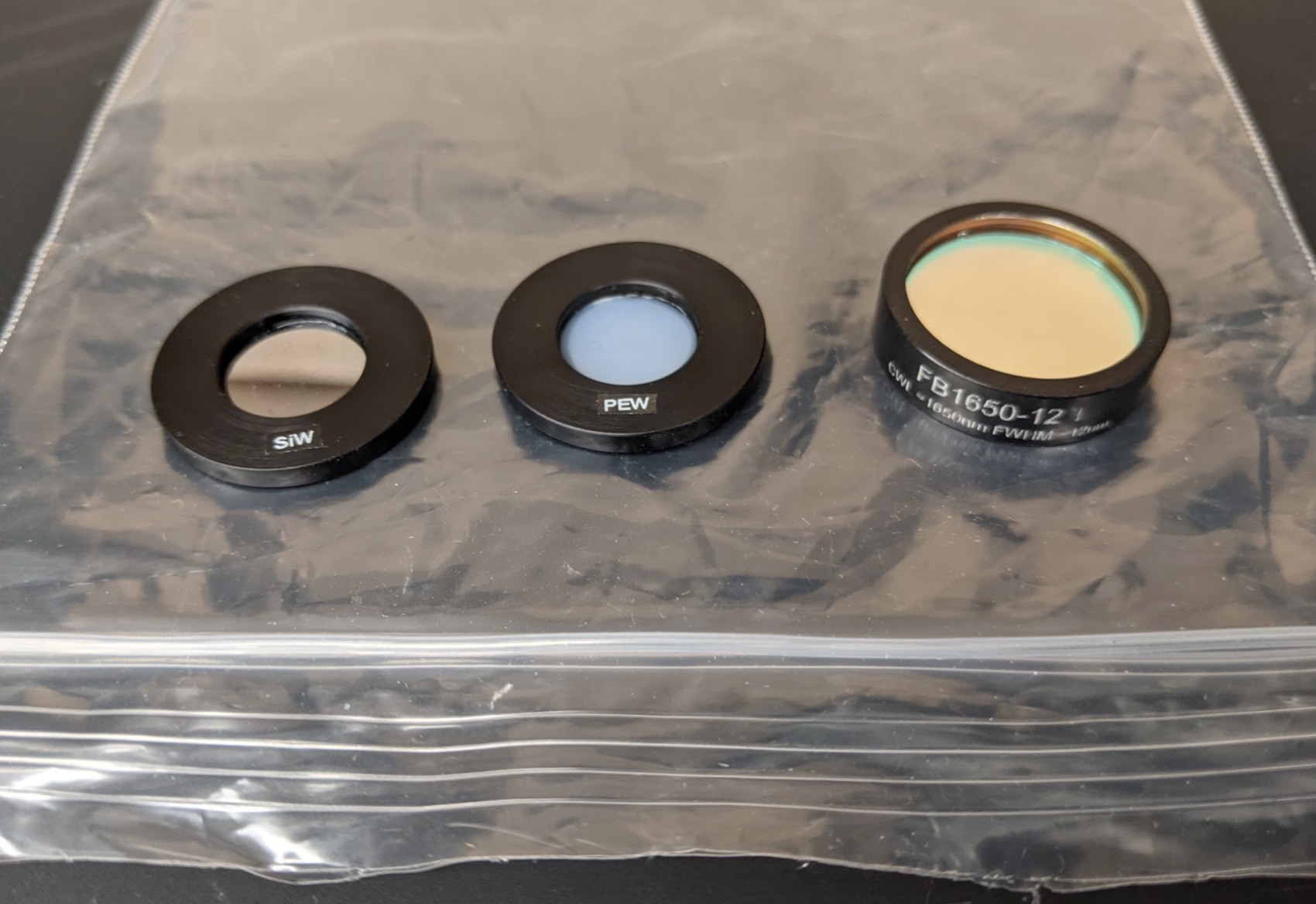}
\caption{\label{fig:filters}
Photograph of optical filters left to right: Gentec SiW (gray), Gentec PEW (light blue), Thorlabs FB1650-12 1650~nm bandpass (gold). 
Each are one inch in diameter, which are directly insert into the detector port. 
These rest on the polyethylene-based plastic bag, which is also considered as a test filter. 
}
\end{figure}

To benchmark the performance of the assembled FTS across an order of magnitude of frequencies, we consider various filters shown in Fig.~\ref{fig:filters}. 
Two filters are obtained from Gentec that are designed for use with the detector: a polyethylene window (PEW) and silicon window (SiW), which have specific absorption characteristics in the mid-infrared.
The manufacturer specifications are 10--1000~$\mu$m (0.3--30 THz) and [1.1--9, 50--1000]~$\mu$m ([0.3--6, 33--270]~THz) design transmission for the PEW and SiW filters, respectively.
These are one inch in diameter and can be directly inserted into the Gentec detector port. 
Additionally, we consider a bandpass filter (Thorlabs FB1650-12), which is designed with center wavelength of $1650\pm 2.4$~nm with $>50\%$ transmission, full-width at half-maximum (FWHM) of $12\pm 2.4$~nm, and sideband rejection  of $<0.01\%$ in the range [200, 1850]~nm. 
We also consider a polyethylene-based resealable plastic bag, as an additional test filter when overlaid over the detector. 

\subsection{\label{sec:operation_analysis}FTS Operation and Analysis}

To align the interferometer optical path, the IR source is temporarily removed and the 635~nm laser is placed directly behind the source location in the optical path. 
We find that the laser is sufficiently bright and the optics performs sufficiently at 635~nm to verify by visual inspection the expected interference fringes at the port of the detector after alignment. 
Once the FTS is aligned and ready for measurements, the IR source is returned to its original position with the emitter centered on the beamspot of the laser. 
This procedure is validated in the interferograms recorded by the detector to be discussed in subsection~\ref{sec:validation_spectra}.
For data processing and visualization of the power measurements, we use the \textsc{NumPy}~\cite{2020NumPy-Array}, \textsc{pandas}~\cite{mckinney-proc-scipy-2010}, and \textsc{matplotlib}~\cite{matplotlib} packages.

The power spectral density $p(f)$ is the Fourier transform $p(f) = \mathcal{F}[p(x)]$ of the input interferogram that measures power as a function of mirror displacement $p(x)$; see the appendix for further discussion.
This is computed numerically using the \texttt{signal.periodogram} routine from \textsc{SciPy}~\cite{2020SciPy-NMeth}, which implements Welch's method~\cite{welch1967use}.
We select the \texttt{parzen} window type after studying the choices available from \texttt{signal.get\_window}~\cite{2020SciPy-NMeth}, based on simulating the interferogram of a square-wave function and examining the signal-to-noise after applying the \texttt{periodogram} routine. 
We define the sampling frequency \texttt{fs} variable in the \texttt{periodogram} routine in terms of instrument parameters by
\begin{equation}
     f_s = 10^{-12} \frac{c R_s }{ 2 v_m} \label{eq:sample_frequency},
\end{equation}
where $c = 3 \times 10^{17}$~nm~s$^{-1}$ is the speed of light, $R_s$ is the Gentec readout sampling frequency, $v_m$ is the mirror scan velocity in nm~s$^{-1}$ assumed to be constant, the $10^{-12}$ factor converts the frequency $f_s$ into units of THz, and the factor of 2 in the denominator accounts for the optical path difference being twice the mirror displacement. 
For a total displacement $L$ (or an integration time of $t=L/v_m$), this defines the theoretical limits of the frequency scan range
\begin{equation}
\frac{c}{2L} < f < \frac{f_s}{2} = \frac{cR_s}{4v_m}.
\end{equation}
For $R_s = 5$~Hz together with typical scan parameters of $v_m = 100$~nm~s$^{-1}$ and $L=0.7$~mm, the corresponding theoretical limits in frequency range is [0.2, 3700]~THz, which is significantly greater than that required for the scope of this paper.
The stage has a maximum travel of $L=12$~mm, which would allow the FTS to achieve a theoretical minimum frequency of 12~GHz.
However, while we expect the mirrors to reflect down to microwave radiation, the absorption of the Gentec photodetector falls substantially for $f \lesssim 10$~THz (Fig.~\ref{fig:GentecAbsorption} in the appendix).
Moreover, we also operate at wavelengths an order of magnitude longer than that specified by the beamsplitter datasheet and performance is expected to be suboptimal. 
Nonetheless, we demonstrate that these optics provide sufficient performance for spectral resolving power down to at least 21~THz in subsection~\ref{sec:filter_transmissions}.

The motorized mirror is controlled by a routine implemented in the Thorlabs \textsc{Kinesis} software, which advances the mirror by the velocity and total displacement set. 
This routine starts (finishes) within two seconds after (before) Gentec data readout begins (ends), where this start-stop synchronization is performed manually and is a current limitation of our setup that could be optimized in future work. 
Nonetheless, this is expected to have a negligible impact within the level of precision of this paper given the typically hour-long duration of an interferogram measurement.
The zero in the $x$-coordinate of mirror displacement is reset in the \textsc{Kinesis} homing routine at the beginning of each day of data taking. 
The mirror displacement is measured relative to this homed zero. 
For the alignment used in this work, we find that at $x = 6.633$~mm, the arms of the Michelson interferometer are equal in length. 
This is identified by the maximum amplitude of the interferogram when using broadband input radiation discussed below, which is referred to as the zero path difference fringe. 
The data are recorded such that zero path difference fringe is located near the central $x$ value of each scan. 

Systematic uncertainties in the instrument parameters of Eq.~\eqref{eq:sample_frequency} affect the measured absolute frequency scale, where uncertainties in $R_s$ and $v_s$ are expected to have the dominant impact.
As an independent cross-check, we directly measure the sampling rate $R_s$ by manually timing the duration for the Gentec readout to record between 1000 to 3000 readings; we find a measured rate $R_s' = 4.81$~Hz, which is smaller than the design 5~Hz. 
Similarly, assuming the displacement reported by the motorized stage is accurate, we also manually timed the duration to travel 0.7~mm using the 100~nm~s$^{-1}$ setting in the \textsc{Kinesis} package and find 119.4 minutes, corresponding to a measured velocity $v_m' = 97.7$~nm~s$^{-1}$. 
Overall, the differences between these direct measurements from the expected correspond to less than 5\%.
However, we can calibrate out these differences and any other instrument effects that impact Eq.~\eqref{eq:sample_frequency}.
This is achieved \emph{in situ} using the bandpass filter centered at 1650~nm and implementing a small correction to the \texttt{fs} variable that centers the measured transmission peak on that expected from the bandpass. 
This data-driven frequency scale calibration is applied to the power spectra to be presented in section~\ref{sec:results}.
The systematic uncertainties after this calibration is expected to be dominated by the instrument resolving power of the 1650~nm lineshape, with a subdominant component from uncertainties of the bandpass specified by the manufacturer.

\section{\label{sec:results}Results and discussion}

Verification that the FTS is successfully aligned to provide expected interferograms and measured power spectra are presented in subsection~\ref{sec:validation_spectra}. 
Potential noise sources are discussed in subsection~\ref{sec:noise_analysis} before subsection~\ref{sec:filter_transmissions} characterizes the FTS performance with measurements of transmission spectra. 

\begin{figure*}[tbh]
\centering
  \includegraphics[width=0.5\linewidth]{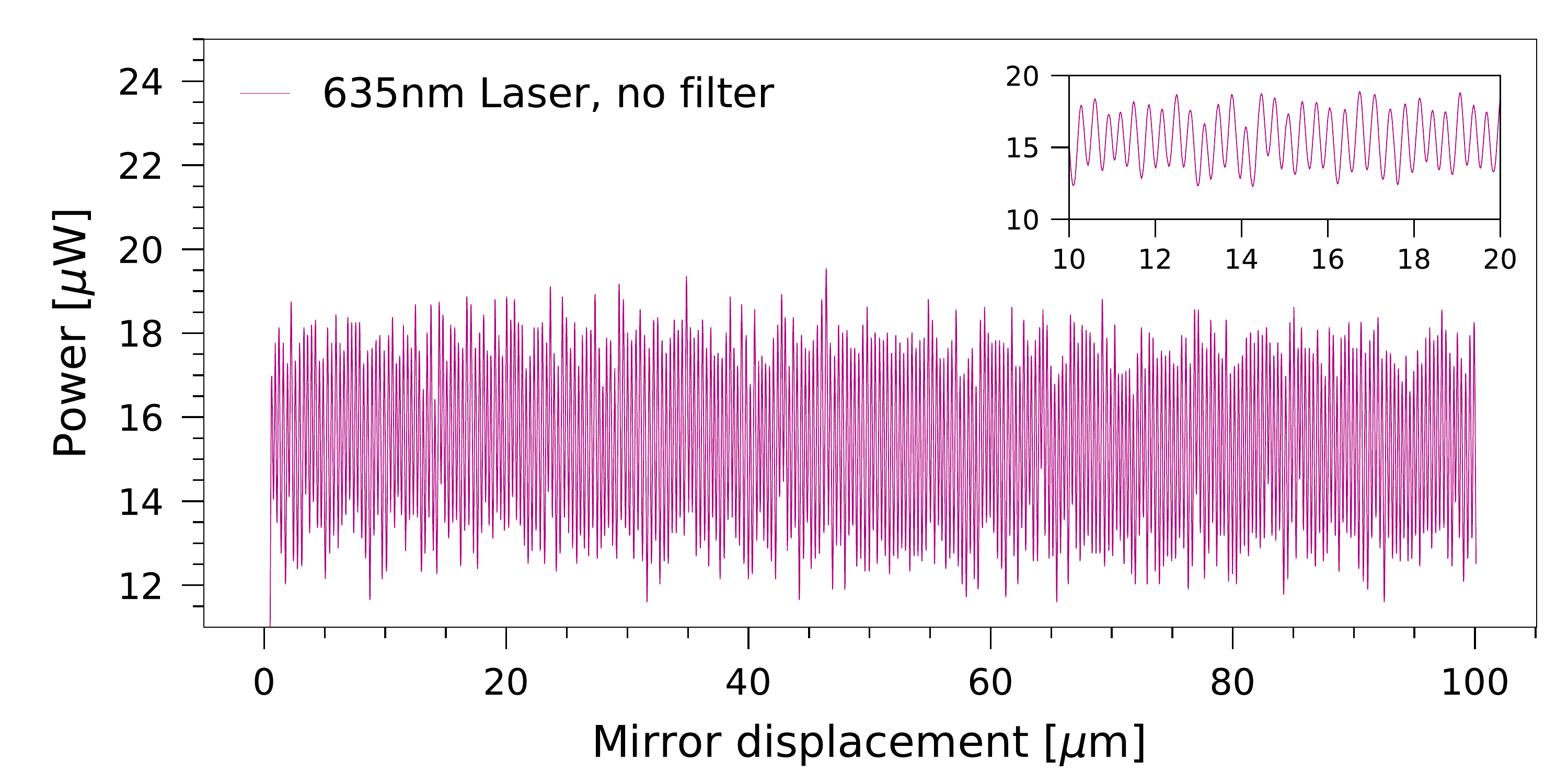}%
  \includegraphics[width=0.5\linewidth]{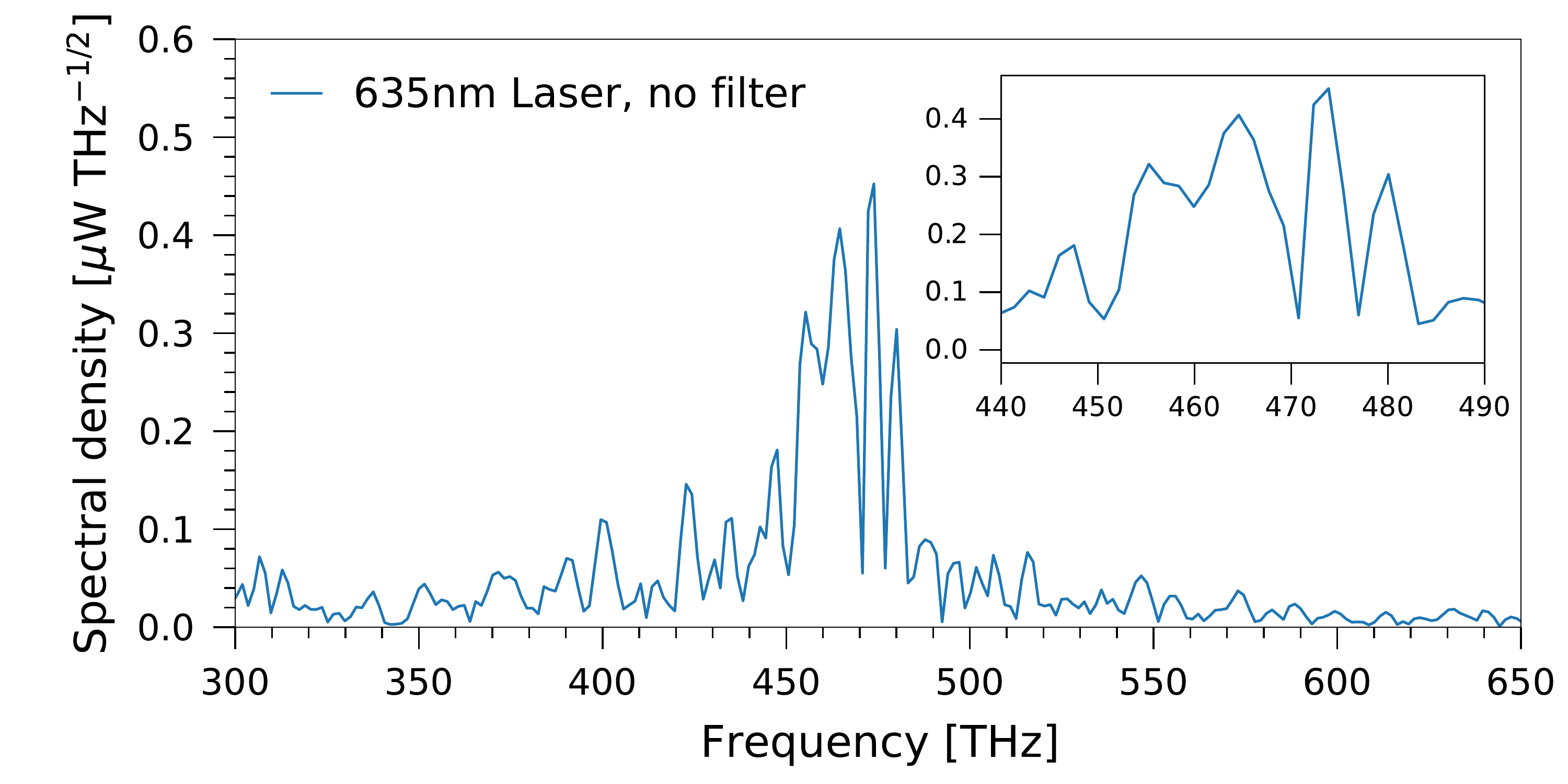}\\
  \includegraphics[width=0.5\linewidth]{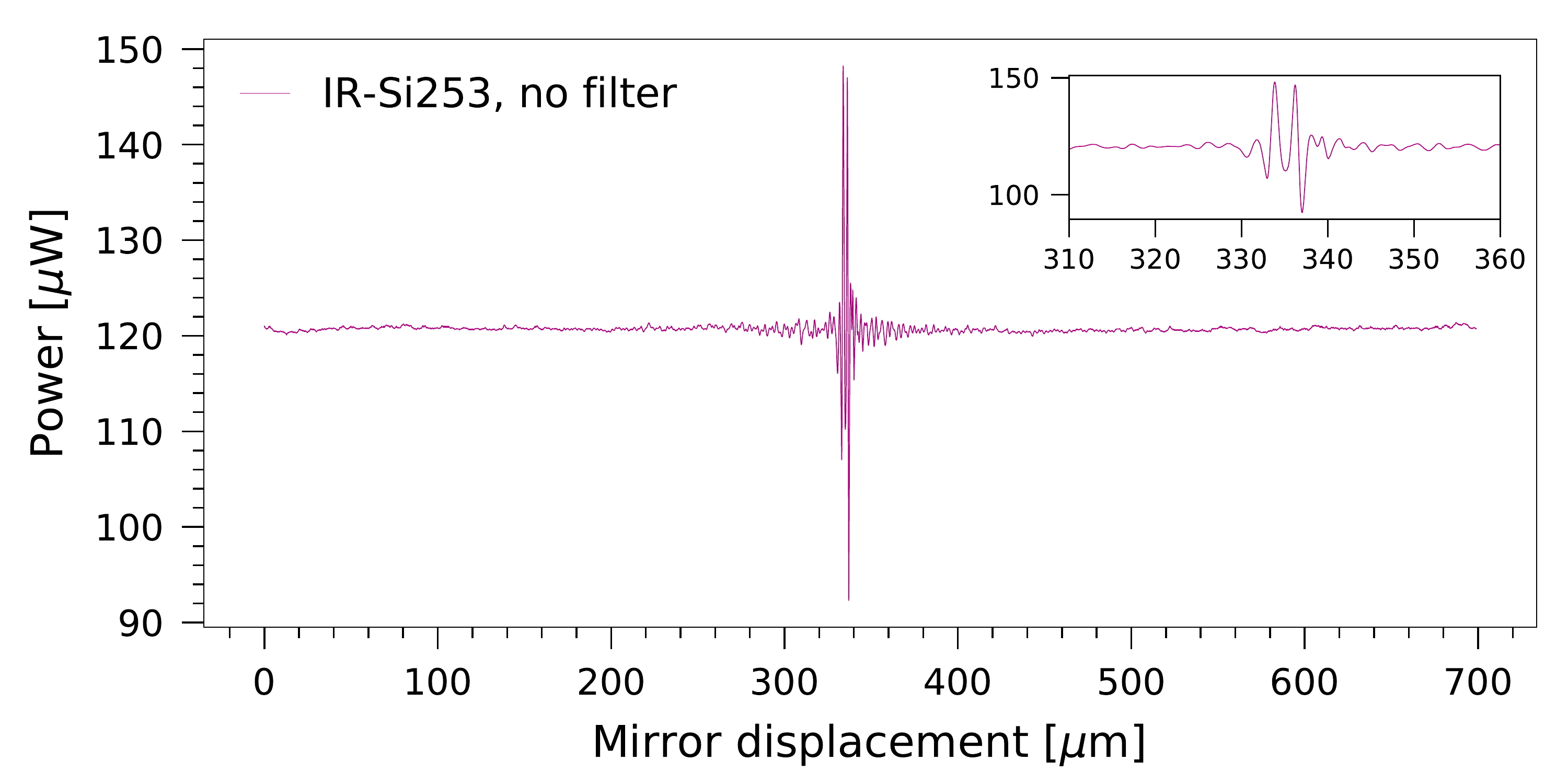}%
  \includegraphics[width=0.5\linewidth]{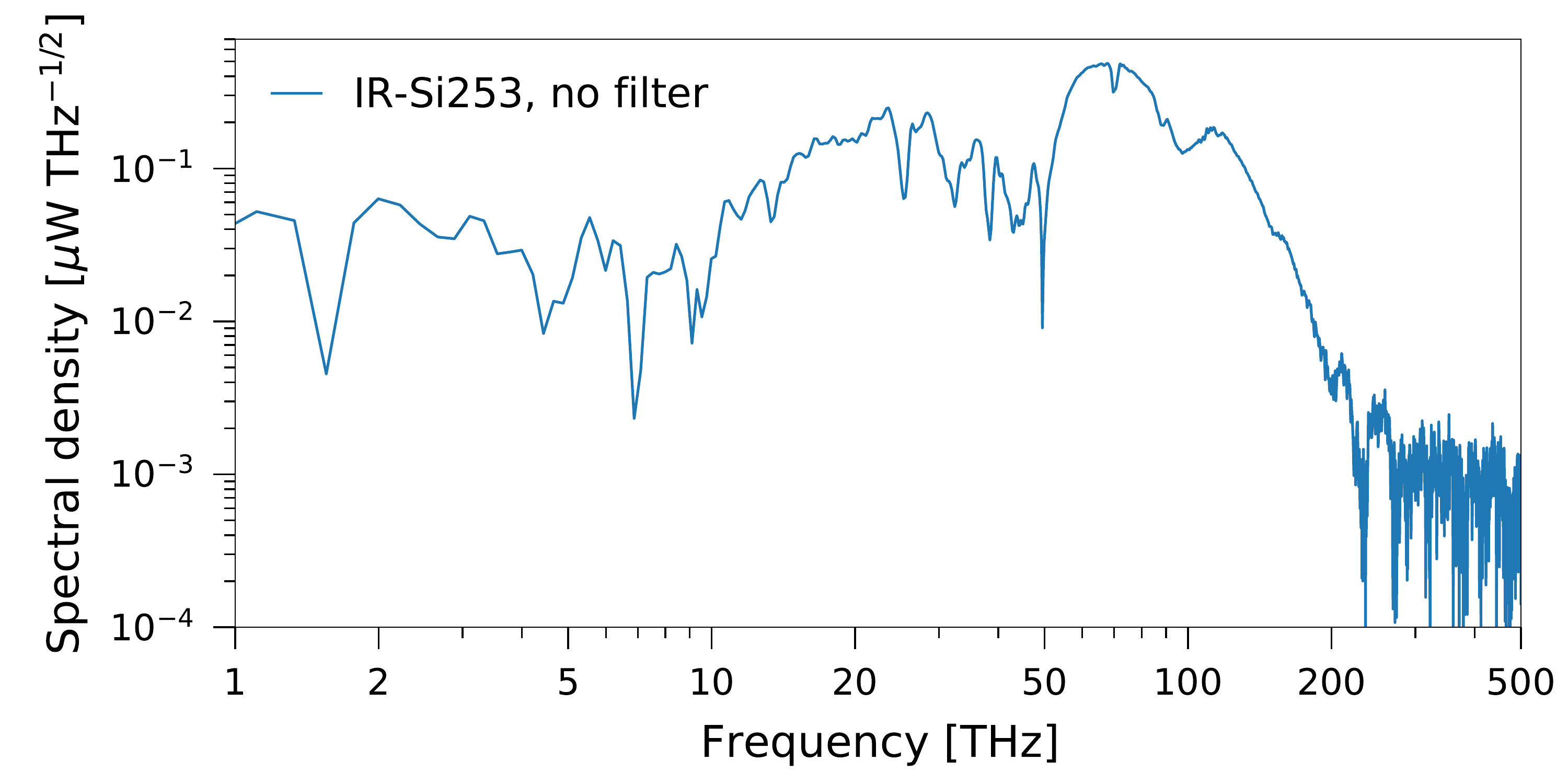}\\
  \includegraphics[width=0.5\linewidth]{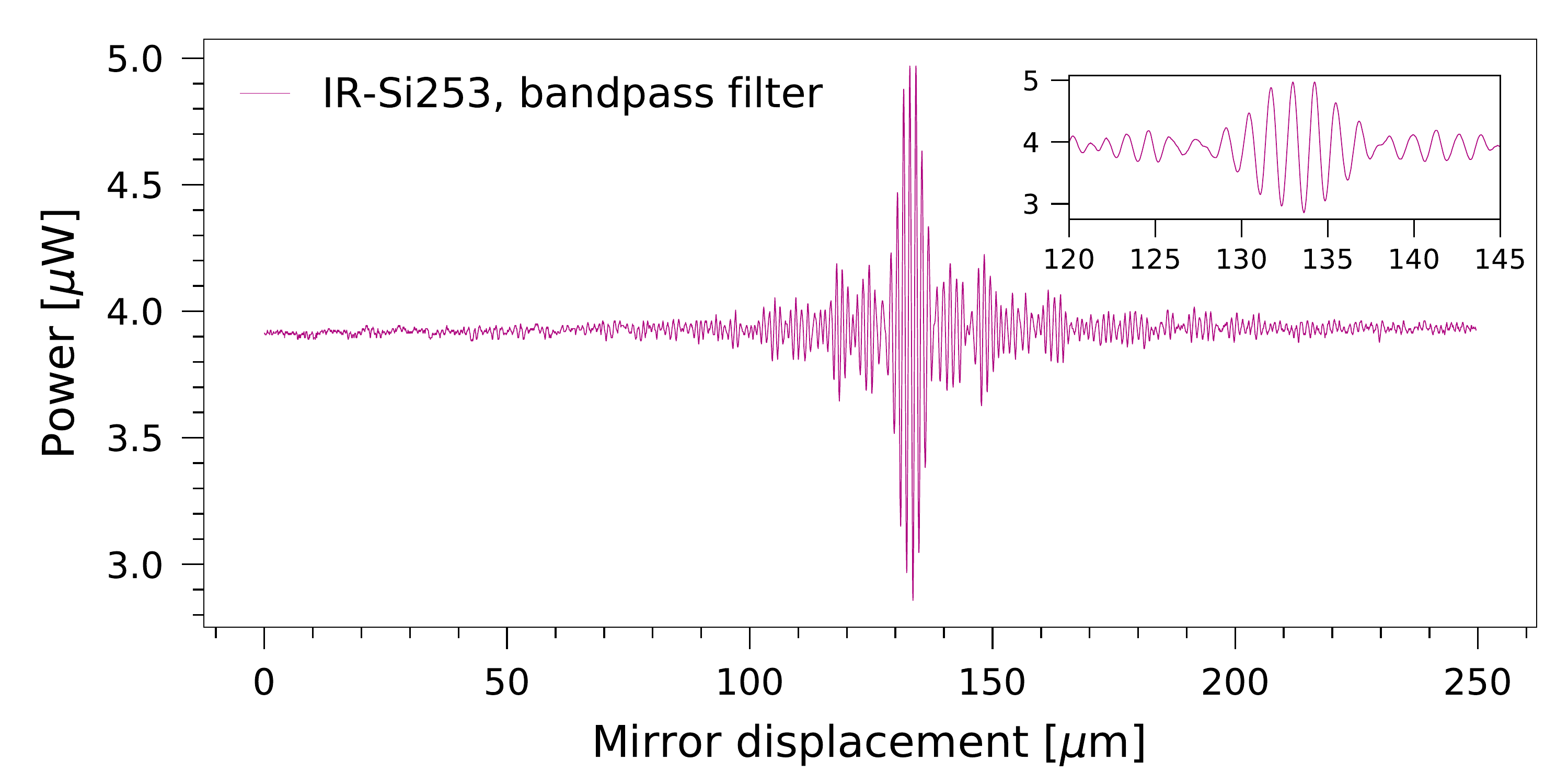}%
  \includegraphics[width=0.5\linewidth]{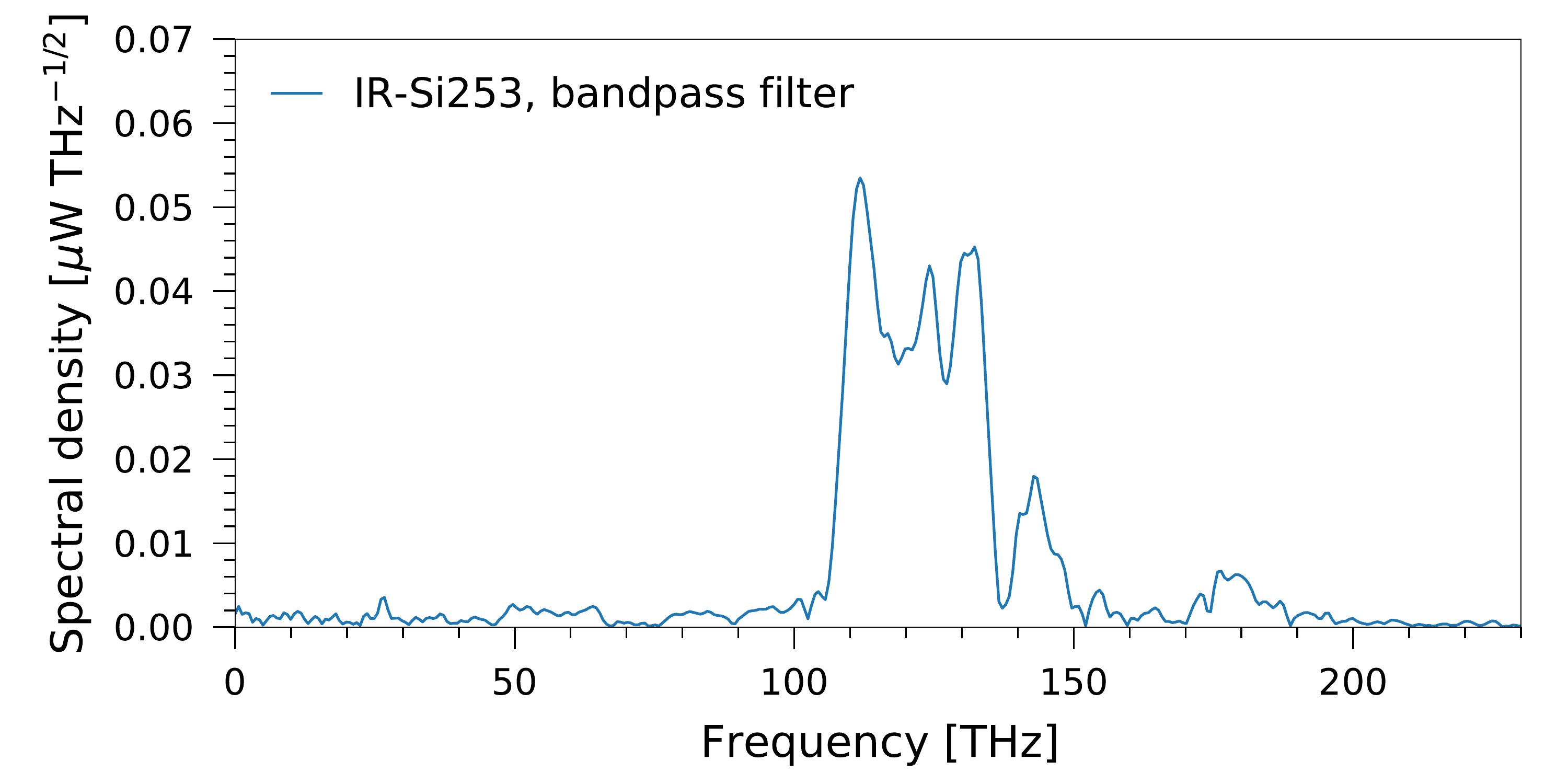}\\
\caption{\label{fig:basic_interferograms}Interferograms of measured power vs mirror displacement (left) and their corresponding power spectral densities after data-driven calibration of frequency scale (right). 
This is displayed for the 635~nm visible laser (top), IR-Si253 source without filters (middle) and with a bandpass filter (bottom). 
The inset plots in the left figures show the same interferograms but zoomed into a smaller range [10, 20]~$\mu$m for the laser, and into the range near zero path difference for the broadband sources. The inset plot on the top right zooms into the laser frequency peak region. }
\end{figure*}

\begin{figure}[h]
    \centering\includegraphics[width=\linewidth]{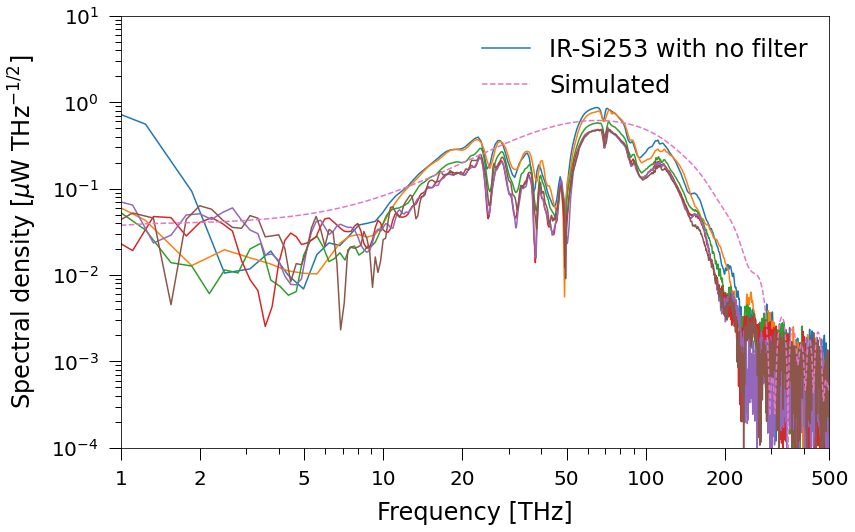}
    \caption{Power spectral density measurements (solid lines) using the IR-Si253 source operating at 7.6~V, 1126~K without filters. 
    The pink dashed line shows the simulation based on Planck's law normalized to that of the six measurements (different colours) for comparison. }
    \label{fig:no_filter_FTS}
\end{figure}

\begin{figure}[h]
\centering
\includegraphics[width=\linewidth]{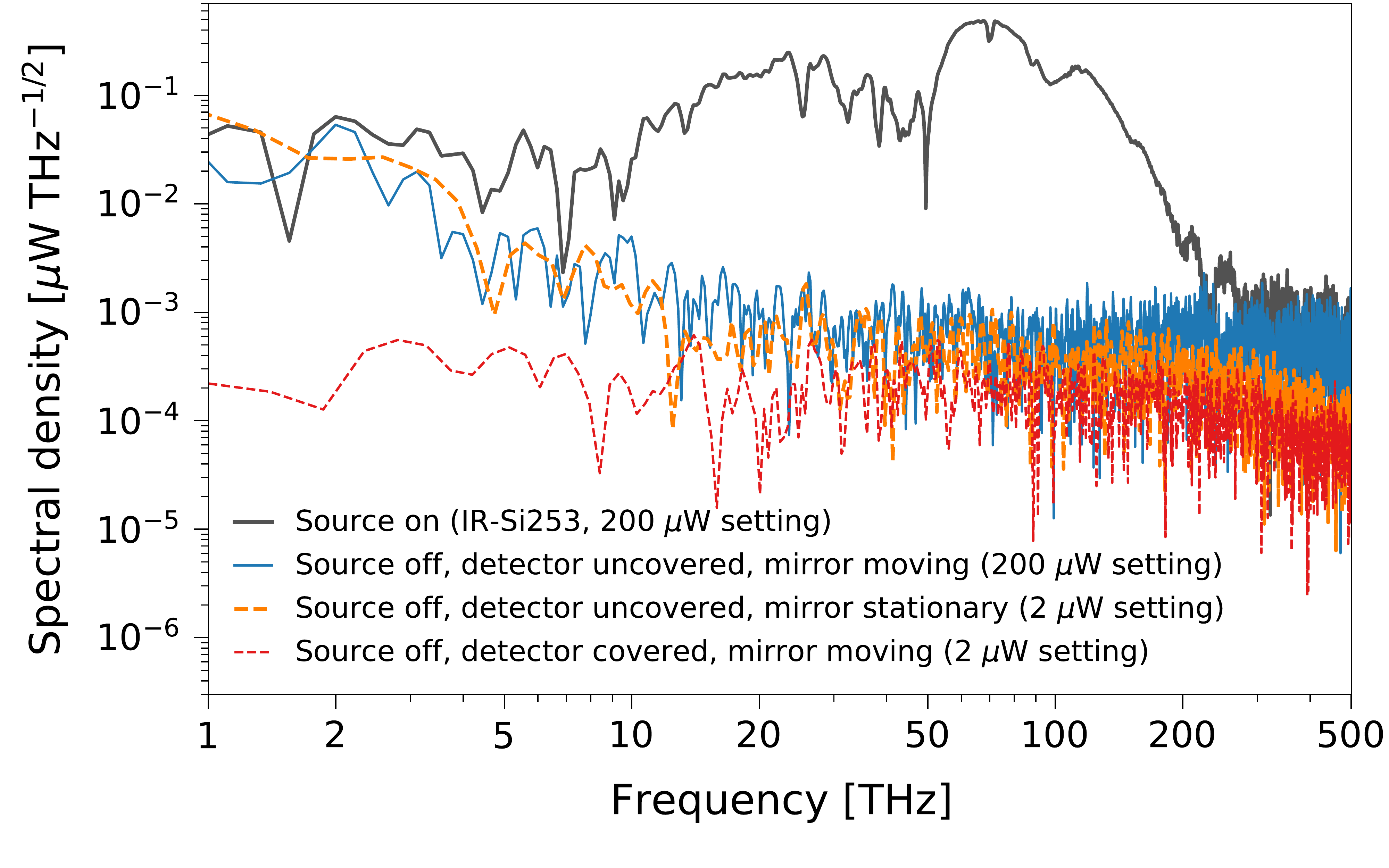}
\caption{\label{fig:noise}Spectrum of the IR-Si253 source (thick solid gray) compared to noise spectra with the source switched off.
The displayed measurements comprise leaving the detector uncovered and moving the mirror at 100~nm~s$^{-1}$ (thin solid blue), then leaving the mirror stationary (thick dashed orange), and finally covering the Gentec detector with the supplied black plastic cover (thin dashed red).
The 200~$\mu$W and 2~$\mu$W settings refer to the power ceiling in the readout with the noise floor being set by the dynamic range of the 12-bit ADC.}
\end{figure}

\subsection{\label{sec:validation_spectra}Validation Spectra}

Using the narrowband 635~nm laser for input radiation, Fig.~\ref{fig:basic_interferograms} (top) shows the recorded interferogram and its Fourier transform (power spectral density). 
The measurements are taken with the motorized stage set to move forward by 100~$\mu$m, with an inset showing [10, 20]~$\mu$m for visual clarity. 
While the optics of the interferometer are designed for the infrared, this laser spectrum verifies that spectroscopy at shorter wavelengths is also technically feasible despite the forthcoming broadband input radiation not being sufficiently bright in the visible to demonstrate this. 
A periodic interferogram with a maximum amplitude of around 3~$\mu$W is measured, which remains relatively constant as the mirror is displaced by 100~$\mu$m. 
The power spectral density in Fig.~\ref{fig:basic_interferograms} (top) is broader than that expected from the laser datasheet and composed of several peaks, where the peak with the highest amplitude is centered around the expected 472~THz position. 
When we directly overlay a pure sinuisoid over the interferogram, we observe beats behaviour. This suggests non-uniformities in the optomechanics that impact the resolution in visible frequencies. 
We could mitigate this by upgrading the stage to one with greater stability or developing real-time synchronization between the readout and stage control, but as our focus is longer wavelengths, this is deferred to future work. 
Nonetheless, this shows that while reconstructing the laser peak in visible frequencies is suboptimal, it is technically feasible.

Turning to broadband IR input radiation, Fig.~\ref{fig:basic_interferograms} (middle) shows the interferogram and the power spectral density for the IR-Si253 source without any designated filters.
The maximum amplitude of the interferogram is found to be 25~$\mu$W, which decays by an order of magnitude after the mirror is displaced 5~$\mu$m away from the zero path difference fringe at 335~$\mu$m, as one would expect for a broadband spectrum. 
The corresponding power spectrum is shown over two orders of magnitude in frequency with many absorption features visible to be discussed further below.

We can compare the general shape of the measured broadband spectra to simulation based on Planck's law for the spectral density of an ideal blackbody radiator
\begin{equation}
B_{f}=\frac{dp}{df d\Omega dA}=\frac{2hf^3}{c^2}\frac{1}{\exp[hf/(k_BT)]-1},
\end{equation}
where $k_{B}$ is Boltzmann's Constant, $h$ is Planck's constant, and $T$ is the temperature of the radiator. 
Figure~\ref{fig:no_filter_FTS} displays the shape predicted by this simulation (solid blue line).
This is compared with the spectra for six measurements with the source fixed at 7.6~V.
The corresponding temperature of 1126~K is determined using the manufacturer datasheet~\cite{BE-IR-Si253}.
There are some differences in absolute power normalization between the six measurements due to small changes in source location, however, measuring features in frequency space are the primary focus of this work rather than absolute power. 
We find reasonable compatibility between the overall shape of the measurements and the expectation from Planck's law, where the expected quadratic growth between 10 and 20~THz and exponential suppression above 150~THz are visible. 
This spectrum forms the no-filter denominator when calculating transmission spectra to be presented in subsection~\ref{sec:filter_transmissions}.
Additionally, several localized absorption features are observed across all six measurements.
These arise from IR-active trace gases in the atmosphere, namely water vapour (H$_2$O) and carbon dioxide (CO$_2$).
The narrow dip at 70~THz corresponds to the CO$_2$ line.
Meanwhile, a set of absorption dips between 25 and 50~THz together with a broad dip around 100~THz are expected from H$_2$O. 
As the IR source is sufficiently bright in the frequency ranges probed in this work, these IR-active gases do not severely impact our transmission measurements.
Nonetheless, future work could mitigate these features by purging the FTS with dry gas such as nitrogen N$_2$.
Equally, that we can identify these expected atmospheric features further verifies the spectral capabilities of our FTS. 

Figure~\ref{fig:basic_interferograms} (bottom) shows the results of this same setup but with the 1650~nm bandpass filter inserted in the detector. 
The interferogram decays more slowly away from zero path difference centered around 130~$\mu$m mirror displacement, corresponding to longer coherence of the radiation. 
Also prominently visible is the periodic modulation of the amplitude (beats) that one would expect when a small finite range of frequencies is selected from broadband radiation.
The power spectral density illustrates the selected frequency range with significant out-of-band rejection below 100~THz frequencies compared with the no-filter spectrum of Fig.~\ref{fig:basic_interferograms} (middle). 
We note that there is leakage between 100 and 160~THz that is expected from the manufacturer datasheet. 
The transmission of this bandpass is discussed below in subsection~\ref{sec:filter_transmissions}.

\subsection{\label{sec:noise_analysis}Noise Analysis}

To characterize noise contributions in the power spectra, the source is switched off and the measurements are displayed in 
Fig.~\ref{fig:noise}. 
We first measure the ambient noise by leaving the detector uncovered with no filters and moving the mirror in a similar manner to data-taking with the source on (solid thin blue). 
The spectral density of this noise rises at lower frequency and has a dependence compatible with $\sim 1/f$.
The signal-to-noise ratio becomes on the order of unity around 5~THz and 200~THz.
The average power of this noise remains below $1~\mu$W and exhibits low-frequency drifts in real space (Fig.~\ref{fig:AmbientNoise} of the appendix).
A similar spectrum is observed when the mirror is required to remain stationary (dashed thick orange).
This suggests that the detector is sensitive to low-frequency time-dependent fluctuations with environmental rather than optomechanical origins. 
We hypothesized that this noise could be due either to time-dependent sensor gain or a time-varying external power source, both of which would violate our assumptions of a stationary signal and generate excess low-frequency noise.
We measured the temperature and humidity at the location of the photodetector using a dedicated USB sensor (Thorlabs TSP01) and found no significant time-dependent correlations with these ambient power drifts.

We next measured the noise with the photosensor obscured using the black plastic cover supplied by Gentec (thin dashed red). 
The spectrum becomes constant as a function of frequency for $\lesssim 100$~THz and is significantly reduced for $\lesssim 20$~THz compared with no-cover measurements, which suggests that the cover decouples the low-frequency time-dependent environmental noise.
This spectrum can be interpreted as the combined noise from the detector and readout, where we observe the integrated power to be less than 0.04~$\mu$W (Fig.~\ref{fig:AmbientNoise} of the appendix), an order of magnitude smaller than the environmental noise. 
The null measurement thus excludes the possibility of purely detector-driven noise. 
The corresponding interferograms in real space for the noise with detector covered as well as uncovered with stationary mirror are displayed in Fig.~\ref{fig:AmbientNoise} of the appendix. 

A final set of tests was performed by moving the chopper from the output of the interferometer to its input, between the beamsplitter and parabolic mirror adjacent to the source.
This excludes environmental sources of input radiation coupling with the detector through the chopper independent of the signal coupled to the interferometer. 
We observe that the measured power falls by an order of magnitude and the noise is compatible with that of detector obscured by the manufacturer-supplied cover. 
This suggests that the non-stationary noise is due to relatively constant ambient power absorbed by the sensor coupled with low-frequency time-dependent drifts from the environment.
These systematic tests ruled out several possible sources of low-frequency noise, but isolating its specific physical origin is deferred to future work.

Furthermore, we find that measurements involving mirror displacements greater than 2~mm induce a noticeable linear growth in integrated power as the mirror is displaced towards the detector. 
This indicates that the beam divergence is significant over longer scan ranges and appears as additional noise artefacts at low frequencies.
Future low-frequency improvements can include null measurements and developing strategies to subtract or correct these low-frequency drifts from the observed spectrum.
Nonetheless, the IR source is sufficiently bright to provide sufficient signal-to-noise for the transmission measurements across the frequency range to be presented in subsection~\ref{sec:filter_transmissions} within the scope of this work.

Finally, at high frequencies $\gtrsim 200$~THz, we find measurements using the 200~$\mu$W setting for the detector power ceiling are bounded by the digitization noise floor of the 12-bit ADC. 
This is shown in Fig.~\ref{fig:noise} by measuring the noise using the lowest 2~$\mu$W setting, where we find that the measured noise is lower at these high frequencies than those measured with the 200~$\mu$W setting.
Future upgrades could consider an ADC with higher dynamic range or dedicated filters and sources more targeted to the frequency range of interest, which will lower the noise floor in future measurements.

\begin{figure}[h]
  \centering
  \includegraphics[width=\linewidth]{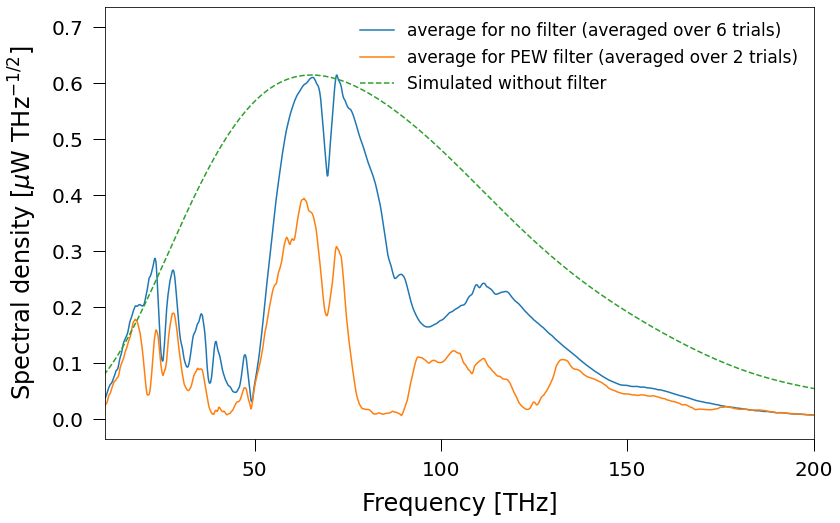}
    \caption{The average of six power spectral measurements in Fig.~\ref{fig:no_filter_FTS} (blue) compared to that with the PEW filter (orange). The expected shape from an ideal blackbody radiator is shown as the dashed green line. 
    }
    \label{fig:PEW_vs_nofilter}
\end{figure}

\subsection{\label{sec:filter_transmissions}Measured Filter Transmission}

\begin{figure}[h]
  \centering
    \includegraphics[width=\linewidth]{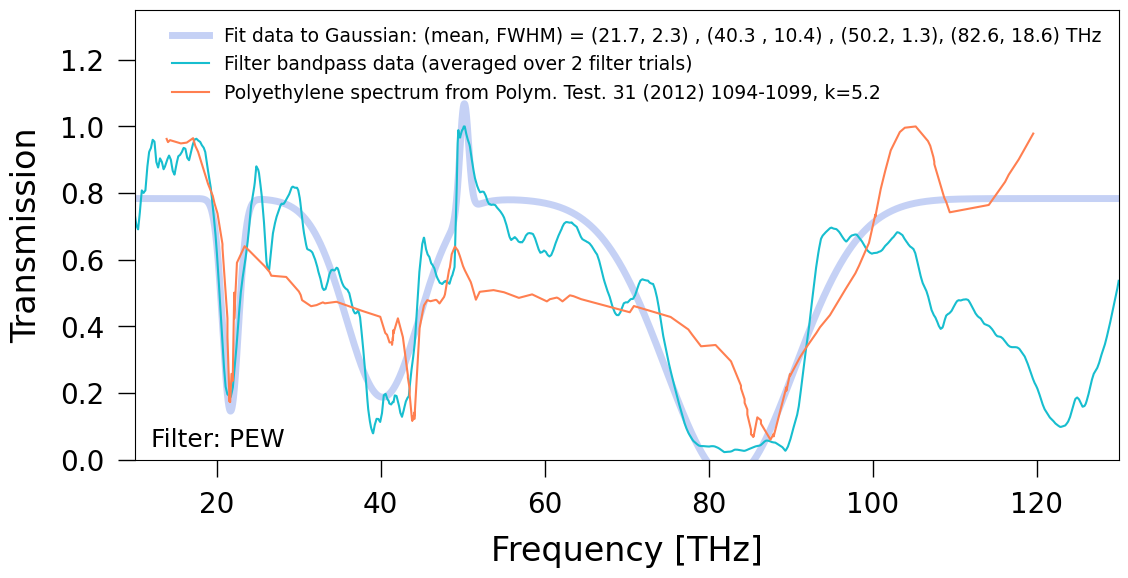}\\
    \includegraphics[width=\linewidth]{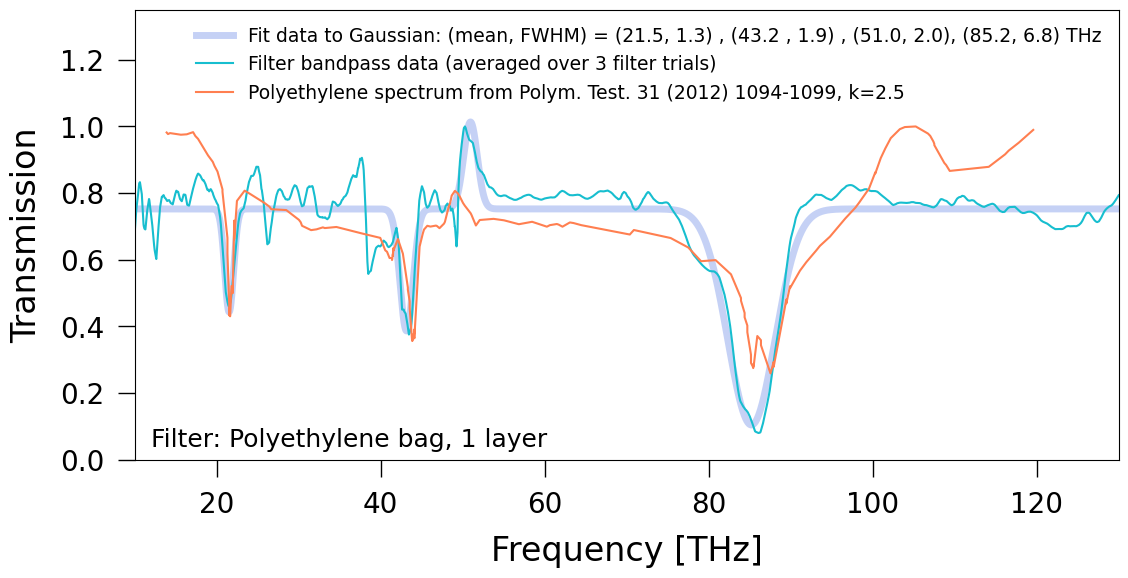}
    \includegraphics[width=\linewidth]{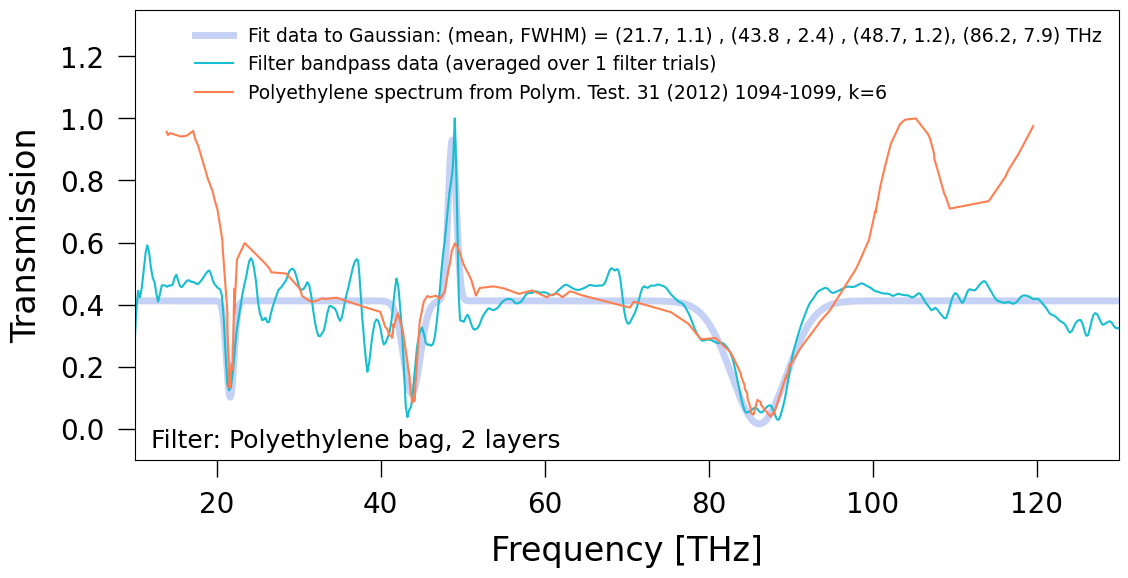}
    \caption{Relative transmission spectra two polyethylene-based filters (cyan) for the PEW filter (top), and polyethylene plastic bag with one layer (middle) and two layers (bottom). This is compared with a polyethylene reference spectrum from Ref.~\cite{PEW_plot} (orange), which has been vertically rescaled to aid comparison of the positions of the spectral features. 
    The thick dark blue line shows a simple model defined in Eq.~\eqref{eq:GaussianFitModel} to  spectral features with the fitted central and full-width at half-maximum values displayed in the legend. }
    \label{fig:filters_vs_polyethylene}
\end{figure}

Figure \ref{fig:PEW_vs_nofilter} shows the arithmetic average of the six measured spectra (blue) shown in Fig.~\ref{fig:no_filter_FTS}, which improves the signal-to-noise.
This is compared with the averaged spectrum when applying the PEW filter (orange), where the overall power is reduced and prominent absorption features are visible. 
The relative transmission of a filter as a function of frequency $T_\text{filter}(f)$ is defined as the ratio of spectra measured with the filter compared to that without
\begin{equation}
    T_\text{filter}(f) = \frac{p_\text{with filter}(f)}{p_\text{no filter}(f)}.
\end{equation}
This is displayed for the PEW filter in Fig.~\ref{fig:filters_vs_polyethylene} (top).
To aid interpretation of the prominent absorption features, we overlay an orange line corresponding to a reference spectrum of polyethylene powder from Ref.~\cite{PEW_plot}, which we rescale vertically to aid visual comparison. 
The physical origin of the narrow features corresponds to molecular modes of the carbon--hydrogen bonds in polyethylene. 
To estimate the properties of our measured spectral features, we construct a simple transmission model $T_\text{fit}$ comprising a linear combination of Gaussian functions $G$
\begin{equation}
    T_\text{fit} = T_C + \sum_i^N G \left(T_i^0, f_i, \delta f_i^\text{FWHM}\right). 
    \label{eq:GaussianFitModel}
\end{equation}
Here, $T_C$ is a free-floating constant offset as a simple description of the sidebands while the Gaussian functions $G$ are parameterized by $T_i^0$ as the height, $f_i$ as the central value, and $\delta f_i^\text{FWHM}$ as the full-width at
half-maximum (FWHM) related to the standard deviation $\sigma$ by $\delta f_i^\text{FWHM} = 2.35 \sigma$. 
The index $i$ labels the $N$ Gaussian functions, where we consider $N=4$ for the polyethylene-based transmission spectra to fit the prominent peaks or dips in data.
The fit (thick blue lines in Fig.~\ref{fig:filters_vs_polyethylene}) is performed by using the least squares method implemented in \texttt{optimize.leastsq} from \textsc{SciPy}~\cite{2020SciPy-NMeth}.

\begin{figure}[h]
  \centering
    \includegraphics[width=\linewidth]{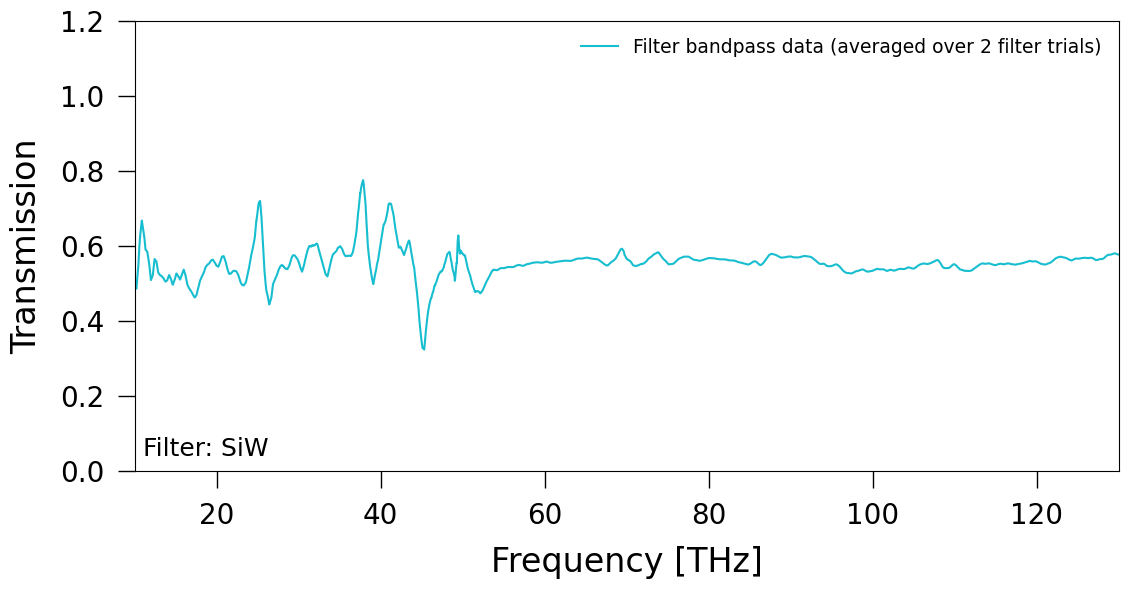}\\
    \includegraphics[width=\linewidth]{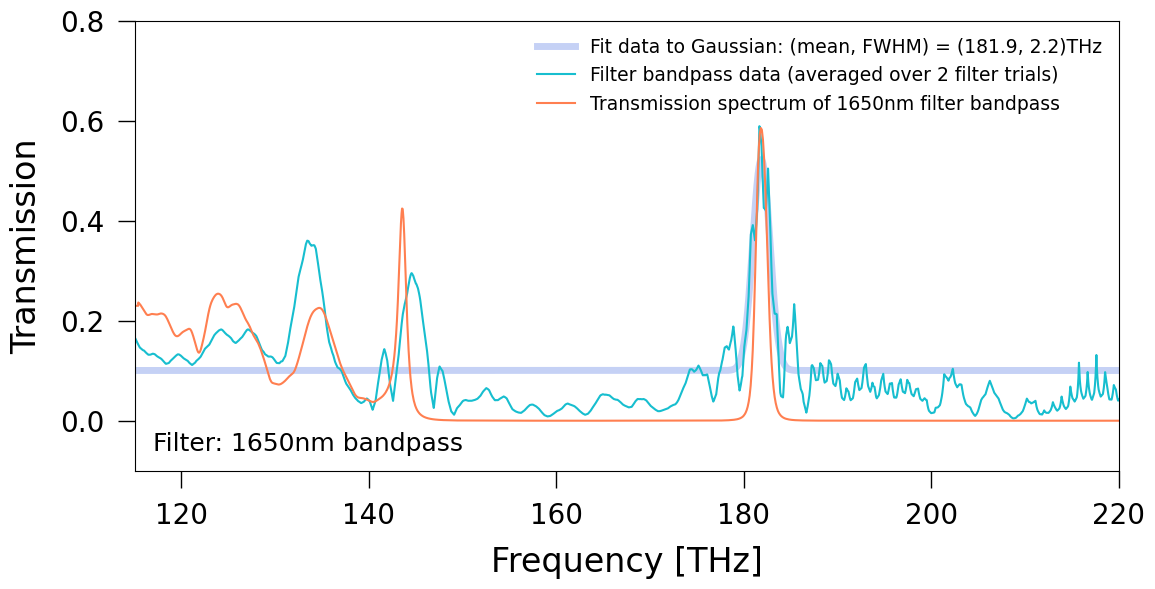}
    \caption{Relative transmission spectra for the SiW silicon window (top), and 1650~nm bandpass filter (bottom). In the bottom plot, the orange line shows the expected transmission of the bandpass from the manufacturer (Thorlabs) datasheet while the thick blue line shows a fit to the bandpass peak. }
    \label{fig:filters_vs_siw_1650nm}
\end{figure}

In the measured PEW transmission, we observe prominent localized absorption lines compatible with those of the reference spectrum at 21~THz, 43~THz, and 87~THz. 
The spectral feature centered at 21~THz has an estimated relative width of $\delta f_\text{FWHM} /f \simeq 10\%$ with a shape that has good compatibility with that of the reference spectrum. 
However, the 43~THz and 87~THz features are notably broader with relative width of $\delta f_\text{FWHM} /f \simeq 27\%$ and $23\%$, respectively, assuming each dip is modelled as one Gaussian. 
These are significantly broader than those in the reference, which suggests a mixed material composition beyond just polyethylene. 
In addition, a narrow feature with high transmission is observed centered around 50~THz with estimated width $\delta f_\text{FWHM} /f \simeq 3\%$.

Figure~\ref{fig:filters_vs_polyethylene} shows the measured transmission of one layer (middle) and two layers (bottom) of the polyethylene plastic bag used as a test filter.  
The measured dip at 21~THz is estimated to be narrower compared with the PEW filter transmission, with a relative width of $\delta f_\text{FWHM} /f \simeq 5.5\%$. 
The absorption features at around 43~THz and 87~THz are narrower compared with that of the PEW filter at $\delta f_\text{FWHM} /f \simeq 6\%$ and 9.4\% respectively, while their shapes are also more compatible with the reference spectrum~\cite{PEW_plot}.
Comparing this plastic bag with the PEW filter above 100~THz, we observe significantly more absorption in the PEW with a minimum around 120~THz. 
Finally, the measured transmission using two layers of the same plastic bag shows around a 40\% decrease in magnitude in the absorption features and sidebands, while the transmission of the 50~THz peak remains as high as the single-layer spectrum.

Overall, these transmission measurements of two polyethylene-based filters demonstrate that our FTS is capable of resolving mid-infrared features with resolution down to at least a few percent.
We find that the measured spectral features are generally compatible with expectations from Ref.~\cite{PEW_plot} despite not expecting exact description given we do not use powdered polyethylene. 
Furthermore, we demonstrate our FTS analysis is capable of distinguishing between similar polyethylene-based materials.
Our analysis suggests that the PEW filter has a richer chemical composition than the polyethylene plastic bag.
This is not unexpected from direct visual inspection (Fig.~\ref{fig:filters}), given the polyethylene bag is non-rigid and transparent to visible light, in contrast to the PEW filter.
However, precise chemical composition analysis as well as interpretation of various secondary spectral features in the sidebands are beyond the scope of this work. 

Figure~\ref{fig:filters_vs_siw_1650nm} (top) shows the transmission for the SiW filter.
The spectrum is qualitatively different from the polyethylene-based filters, as one would expect from its inorganic composition, further underscoring the ability of our FTS to discriminate between materials. 
A relatively uniform transmission is observed around 50\% above 50~THz along with localized peaks below 50~THz. We find little correlation between the measured transmission spectra of the PEW and SiW filters with the manufacturer-specified frequency ranges.
Figure~\ref{fig:filters_vs_siw_1650nm} (bottom) shows the transmission of the 1650~nm bandpass filter. 
This benchmarks our FTS at near-IR frequencies with a known transmission profile specified by the manufacturer datasheet.
Our measurement gives $\delta f_\text{FWHM} /f \simeq 1.2\%$, compared with the datasheet width of $\delta f_\text{FWHM} /f \simeq 0.73\%$, providing an indication of our FTS spectral resolution.
We observe modest transmission below 150~THz that is expected from the datasheet; we note that out-of-band blocking is only specified by the manufacturer above 162~THz (1850~nm). 

\section{\label{sec:summary}Summary and future directions}

In summary, we have constructed a tabletop Fourier transform spectrometer using commercial optical components designed for mid-infrared frequencies arranged as a Michelson interferometer. 
This is coupled to a broadband pyroelectric photodetector with design sensitivity to [0.1, 30]~THz frequencies. 
After successful alignment using a visible laser,
we used a silicon nitride infrared source to provide broadband input radiation to characterize the instrument performance over an order of magnitude in frequency [10, 200]~THz, where the source is sufficiently bright to yield significant signal-to-noise. 
We used a bandpass filter in frequencies that the mid-infrared optics are designed for to demonstrate the reconstruction of a narrow line at 182~THz (1650~nm), and used this for \emph{in situ} calibration of the frequency scale.
Using polyethylene-based filters, 
we demonstrated performance at frequencies lower than those specified by the mid-infrared optics, beginning to probe the spectral regime where the pyroelectric detector is designed for ($<30$~THz).  
Specifically, we identified multiple spectral lines between 20 and 130~THz with compatibility between a dedicated filter, a polyethylene-based plastic bag, and expectation from a reference powder spectrum in the literature.
This enables discrimination between filters composed of different materials.

In future work, we plan to extend the spectral analysis capabilities of our FTS below 10~THz.
We expect the current optomechanics of our instrument to be sufficient.
However, measurements below 10~THz are currently limited by noise in part due to the precipitous drop in low-frequency source power while the bright IR peak around 70~THz saturates the dynamic range of the detector readout.  
This could be resolved using a brighter source coupled to a low-pass filter to reject the IR peak, while developing strategies to suppress environmental and instrument noise contributing to low-frequency regimes that we identified. 
Specifically, we could mitigate non-stationary noise through the addition of a second optical chopper at the FTS input. 
We could also test dedicated optics designed for this terahertz region rather than relying on optics designed for the mid-infrared.
A current operational limitation of our photodetector is the low sampling frequency of 5~Hz, leading to modest data-taking duration of two hours for 0.7~mm mirror displacements. 
We could also synchronize the detector readout and motorized stage control in real-time rather than assuming uniform motion. 
Future work could test our interferometer with more advanced sources, such as terahertz quantum cascade lasers.
Significant noise reduction could be possible by extending the FTS design to cryogenic operation and coupling to superconducting kinetic inductance detectors for low-noise sensitivity at lower frequencies. 

Longer term, we anticipate investigating applications of a dedicated terahertz FTS in proposed axion dark matter experiments~\cite{THzAxionSnowmassLoI,THzAxionSonnenschein}.
The FTS is appealing as it is broadband by design with good spectral resolution, which is aligned with recent interests in astroparticle physics to develop broadband search techniques for the poorly constrained [0.1, 100]~meV DM mass range, corresponding to frequencies we studied. 
Broadband search capability is important given the DM mass is \emph{a priori} unknown, but there are theoretical and cosmological arguments that motivate this mass region. 
We expect our FTS can also be used to characterize dedicated optics in these frequency ranges before use in the DM experiment for calibration. 
A further detection challenge is the small coupling between DM and photons, leading to low signal rates. 
Sensitivity therefore requires significant noise suppression via cryogenic operation and use of superconducting single-photon counting detectors. 
Coupling an FTS as a spectral analyzer to these experiments could further enhance detection sensitivity because the DM induces monochromatic photon emission that can be resolved as a spectral peak above noise.
After DM is discovered, this capability enables direct mass measurement given it is proportional to the photon frequency and demands good spectral resolution. 
These opportunities motivate planned upgrades of our instrument to improve spectral range, resolution, and noise suppression.
Our initial FTS setup paves the way to interesting interdisciplinary applications of multi-terahertz spectrometers in dark matter physics and beyond.

\emph{\textbf{Acknowledgements}}---We are grateful to Ankur Agrawal, Adam Anderson, Kelby Anderson, Daniel Bowring, Aaron Chou, Amanda Farah, Casey Frantz, Mary Heintz, Rakshya Khatiwada, Jan Offermann, Mark Oreglia, Omid Noroozian, Jessica Schmidt, and Emily Smith for laboratory assistance and helpful discussions. 
This work is funded in part by the Department of Energy through the program for Quantum Information Science Enabled Discovery (QuantISED) for High Energy Physics and the resources of the Fermi National Accelerator Laboratory (Fermilab), a U.S. Department of Energy, Office of Science, HEP User Facility. Fermilab is managed by Fermi Research Alliance, LLC (FRA), acting under Contract No. DE-AC02-07CH11359.
Work at Argonne National Laboratory is supported by the U.S. Department of Energy, Office of High Energy Physics, under contract DE-AC02-06CH11357.
We acknowledge support by the Kavli Institute for Cosmological Physics at the University of Chicago through grant NSF PHY-1125897 and an endowment from the Kavli Foundation and its founder Fred Kavli.
JL is supported by the Grainger Fellowship at the University of Chicago. 

\bibliography{refs}

\begin{thebibliography}{69}%
\makeatletter
\providecommand \@ifxundefined [1]{%
 \@ifx{#1\undefined}
}%
\providecommand \@ifnum [1]{%
 \ifnum #1\expandafter \@firstoftwo
 \else \expandafter \@secondoftwo
 \fi
}%
\providecommand \@ifx [1]{%
 \ifx #1\expandafter \@firstoftwo
 \else \expandafter \@secondoftwo
 \fi
}%
\providecommand \natexlab [1]{#1}%
\providecommand \enquote  [1]{``#1''}%
\providecommand \bibnamefont  [1]{#1}%
\providecommand \bibfnamefont [1]{#1}%
\providecommand \citenamefont [1]{#1}%
\providecommand \href@noop [0]{\@secondoftwo}%
\providecommand \href [0]{\begingroup \@sanitize@url \@href}%
\providecommand \@href[1]{\@@startlink{#1}\@@href}%
\providecommand \@@href[1]{\endgroup#1\@@endlink}%
\providecommand \@sanitize@url [0]{\catcode `\\12\catcode `\$12\catcode
  `\&12\catcode `\#12\catcode `\^12\catcode `\_12\catcode `\%12\relax}%
\providecommand \@@startlink[1]{}%
\providecommand \@@endlink[0]{}%
\providecommand \url  [0]{\begingroup\@sanitize@url \@url }%
\providecommand \@url [1]{\endgroup\@href {#1}{\urlprefix }}%
\providecommand \urlprefix  [0]{URL }%
\providecommand \Eprint [0]{\href }%
\providecommand \doibase [0]{https://doi.org/}%
\providecommand \selectlanguage [0]{\@gobble}%
\providecommand \bibinfo  [0]{\@secondoftwo}%
\providecommand \bibfield  [0]{\@secondoftwo}%
\providecommand \translation [1]{[#1]}%
\providecommand \BibitemOpen [0]{}%
\providecommand \bibitemStop [0]{}%
\providecommand \bibitemNoStop [0]{.\EOS\space}%
\providecommand \EOS [0]{\spacefactor3000\relax}%
\providecommand \BibitemShut  [1]{\csname bibitem#1\endcsname}%
\let\auto@bib@innerbib\@empty
\bibitem [{\citenamefont {Griffiths}\ \emph {et~al.}(2007)\citenamefont
  {Griffiths}, \citenamefont {De~Haseth},\ and\ \citenamefont
  {Winefordner}}]{griffiths2007fourier}%
  \BibitemOpen
  \bibfield  {author} {\bibinfo {author} {\bibfnamefont {P.}~\bibnamefont
  {Griffiths}}, \bibinfo {author} {\bibfnamefont {J.}~\bibnamefont
  {De~Haseth}},\ and\ \bibinfo {author} {\bibfnamefont {J.}~\bibnamefont
  {Winefordner}},\ }\href {https://books.google.com/books?id=C\_c0GVe8MX0C}
  {\emph {\bibinfo {title} {Fourier Transform Infrared Spectrometry}}},\
  Chemical Analysis: A Series of Monographs on Analytical Chemistry and Its
  Applications\ (\bibinfo  {publisher} {Wiley},\ \bibinfo {year}
  {2007})\BibitemShut {NoStop}%
\bibitem [{\citenamefont {{Newport Corporation}}()}]{NewportFTIR}%
  \BibitemOpen
  \bibfield  {author} {\bibinfo {author} {\bibnamefont {{Newport
  Corporation}}},\ }\href@noop {} {}\bibinfo {note}
  {\href{https://www.newport.com/n/introduction-to-ftir-spectroscopy}{Technical
  Note: Introduction to FTIR Spectroscopy}}\BibitemShut {NoStop}%
\bibitem [{\citenamefont {Movasaghi}\ \emph {et~al.}(2008)\citenamefont
  {Movasaghi}, \citenamefont {Rehman},\ and\ \citenamefont
  {ur~Rehman}}]{Zanyar2008}%
  \BibitemOpen
  \bibfield  {author} {\bibinfo {author} {\bibfnamefont {Z.}~\bibnamefont
  {Movasaghi}}, \bibinfo {author} {\bibfnamefont {S.}~\bibnamefont {Rehman}},\
  and\ \bibinfo {author} {\bibfnamefont {D.~I.}\ \bibnamefont {ur~Rehman}},\
  }\bibfield  {title} {\bibinfo {title} {{Fourier Transform Infrared (FTIR)
  Spectroscopy of Biological Tissues}},\ }\href
  {https://doi.org/10.1080/05704920701829043} {\bibfield  {journal} {\bibinfo
  {journal} {Appl. Spectrosc. Rev.}\ }\textbf {\bibinfo {volume} {43}},\
  \bibinfo {pages} {134} (\bibinfo {year} {2008})}\BibitemShut {NoStop}%
\bibitem [{\citenamefont {Geibel}\ \emph {et~al.}(2010)\citenamefont {Geibel},
  \citenamefont {Gerbig},\ and\ \citenamefont {Feist}}]{geibel2010new}%
  \BibitemOpen
  \bibfield  {author} {\bibinfo {author} {\bibfnamefont {M.~C.}\ \bibnamefont
  {Geibel}}, \bibinfo {author} {\bibfnamefont {C.}~\bibnamefont {Gerbig}},\
  and\ \bibinfo {author} {\bibfnamefont {D.~G.}\ \bibnamefont {Feist}},\
  }\bibfield  {title} {\bibinfo {title} {{A new fully automated FTIR system for
  total column measurements of greenhouse gases}},\ }\href
  {https://doi.org/10.5194/amt-3-1363-2010} {\bibfield  {journal} {\bibinfo
  {journal} {Atmos. Meas. Tech.}\ }\textbf {\bibinfo {volume} {3}},\ \bibinfo
  {pages} {1363} (\bibinfo {year} {2010})}\BibitemShut {NoStop}%
\bibitem [{\citenamefont {Bellisola}\ and\ \citenamefont
  {Sorio}(2012)}]{bellisola2012infrared}%
  \BibitemOpen
  \bibfield  {author} {\bibinfo {author} {\bibfnamefont {G.}~\bibnamefont
  {Bellisola}}\ and\ \bibinfo {author} {\bibfnamefont {C.}~\bibnamefont
  {Sorio}},\ }\bibfield  {title} {\bibinfo {title} {{Infrared spectroscopy and
  microscopy in cancer research and diagnosis}},\ }\href
  {https://europepmc.org/article/pmc/pmc3236568} {\bibfield  {journal}
  {\bibinfo  {journal} {Am. J. Cancer Res.}\ }\textbf {\bibinfo {volume} {2}},\
  \bibinfo {pages} {1} (\bibinfo {year} {2012})}\BibitemShut {NoStop}%
\bibitem [{\citenamefont {Baker}\ \emph {et~al.}(2014)\citenamefont {Baker}
  \emph {et~al.}}]{baker2014using}%
  \BibitemOpen
  \bibfield  {author} {\bibinfo {author} {\bibfnamefont {M.~J.}\ \bibnamefont
  {Baker}} \emph {et~al.},\ }\bibfield  {title} {\bibinfo {title} {{Using
  Fourier transform IR spectroscopy to analyze biological materials}},\ }\href
  {https://doi.org/10.1038/nprot.2014.110} {\bibfield  {journal} {\bibinfo
  {journal} {Nat. Protoc.}\ }\textbf {\bibinfo {volume} {9}},\ \bibinfo {pages}
  {1771} (\bibinfo {year} {2014})}\BibitemShut {NoStop}%
\bibitem [{\citenamefont {Gisi}\ \emph {et~al.}(2012)\citenamefont {Gisi},
  \citenamefont {Hase}, \citenamefont {Dohe}, \citenamefont {Blumenstock},
  \citenamefont {Simon},\ and\ \citenamefont {Keens}}]{gisi2012xco2}%
  \BibitemOpen
  \bibfield  {author} {\bibinfo {author} {\bibfnamefont {M.}~\bibnamefont
  {Gisi}}, \bibinfo {author} {\bibfnamefont {F.}~\bibnamefont {Hase}}, \bibinfo
  {author} {\bibfnamefont {S.}~\bibnamefont {Dohe}}, \bibinfo {author}
  {\bibfnamefont {T.}~\bibnamefont {Blumenstock}}, \bibinfo {author}
  {\bibfnamefont {A.}~\bibnamefont {Simon}},\ and\ \bibinfo {author}
  {\bibfnamefont {A.}~\bibnamefont {Keens}},\ }\bibfield  {title} {\bibinfo
  {title} {{XCO$_2$-measurements with a tabletop FTS using solar absorption
  spectroscopy}},\ }\href {https://doi.org/10.5194/amt-5-2969-2012} {\bibfield
  {journal} {\bibinfo  {journal} {Atmos. Meas. Tech.}\ }\textbf {\bibinfo
  {volume} {5}},\ \bibinfo {pages} {2969} (\bibinfo {year} {2012})}\BibitemShut
  {NoStop}%
\bibitem [{\citenamefont {Cossel}\ \emph {et~al.}(2017)\citenamefont {Cossel},
  \citenamefont {Waxman}, \citenamefont {Finneran}, \citenamefont {Blake},
  \citenamefont {Ye},\ and\ \citenamefont {Newbury}}]{cossel2017gas}%
  \BibitemOpen
  \bibfield  {author} {\bibinfo {author} {\bibfnamefont {K.~C.}\ \bibnamefont
  {Cossel}}, \bibinfo {author} {\bibfnamefont {E.~M.}\ \bibnamefont {Waxman}},
  \bibinfo {author} {\bibfnamefont {I.~A.}\ \bibnamefont {Finneran}}, \bibinfo
  {author} {\bibfnamefont {G.~A.}\ \bibnamefont {Blake}}, \bibinfo {author}
  {\bibfnamefont {J.}~\bibnamefont {Ye}},\ and\ \bibinfo {author}
  {\bibfnamefont {N.~R.}\ \bibnamefont {Newbury}},\ }\bibfield  {title}
  {\bibinfo {title} {Gas-phase broadband spectroscopy using active sources:
  progress, status, and applications},\ }\href
  {https://doi.org/10.1364/JOSAB.34.000104} {\bibfield  {journal} {\bibinfo
  {journal} {J. Opt. Soc. Am. B}\ }\textbf {\bibinfo {volume} {34}},\ \bibinfo
  {pages} {104} (\bibinfo {year} {2017})}\BibitemShut {NoStop}%
\bibitem [{\citenamefont {Mather}\ \emph {et~al.}(1993)\citenamefont {Mather},
  \citenamefont {Fixsen},\ and\ \citenamefont {Shafer}}]{Mather1993}%
  \BibitemOpen
  \bibfield  {author} {\bibinfo {author} {\bibfnamefont {J.~C.}\ \bibnamefont
  {Mather}}, \bibinfo {author} {\bibfnamefont {D.~J.}\ \bibnamefont {Fixsen}},\
  and\ \bibinfo {author} {\bibfnamefont {R.~A.}\ \bibnamefont {Shafer}},\
  }\bibfield  {title} {\bibinfo {title} {{Design for the COBE far-infrared
  absolute spectrophotometer (FIRAS)}},\ }in\ \href
  {https://doi.org/10.1117/12.157823} {\emph {\bibinfo {booktitle} {Infrared
  Spaceborne Remote Sensing}}},\ Vol.\ \bibinfo {volume} {2019},\ \bibinfo
  {editor} {edited by\ \bibinfo {editor} {\bibfnamefont {M.~S.}\ \bibnamefont
  {Scholl}}},\ \bibinfo {organization} {International Society for Optics and
  Photonics}\ (\bibinfo  {publisher} {SPIE},\ \bibinfo {year} {1993})\ pp.\
  \bibinfo {pages} {168 -- 179}\BibitemShut {NoStop}%
\bibitem [{\citenamefont {N{\ae}ss}\ \emph {et~al.}(2019)\citenamefont
  {N{\ae}ss}, \citenamefont {Dunkley}, \citenamefont {Kogut},\ and\
  \citenamefont {Fixsen}}]{naess2019time}%
  \BibitemOpen
  \bibfield  {author} {\bibinfo {author} {\bibfnamefont {S.~K.}\ \bibnamefont
  {N{\ae}ss}}, \bibinfo {author} {\bibfnamefont {J.}~\bibnamefont {Dunkley}},
  \bibinfo {author} {\bibfnamefont {A.}~\bibnamefont {Kogut}},\ and\ \bibinfo
  {author} {\bibfnamefont {D.~J.}\ \bibnamefont {Fixsen}},\ }\bibfield  {title}
  {\bibinfo {title} {{Time-ordered data simulation and map-making for the PIXIE
  Fourier transform spectrometer}},\ }\href
  {https://doi.org/10.1088/1475-7516/2019/04/019} {\bibfield  {journal}
  {\bibinfo  {journal} {J. Cosmol. Astropart. Phys.}\ }\textbf {\bibinfo
  {volume} {2019}}\bibfield  {number} {\bibinfo  {number} { (04)},\ \bibinfo
  {pages} {019}},\ }\Eprint {https://arxiv.org/abs/1710.06761}
  {arXiv:1710.06761} \BibitemShut {NoStop}%
\bibitem [{\citenamefont {Carlstrom}\ \emph {et~al.}(2011)\citenamefont
  {Carlstrom} \emph {et~al.}}]{carlstrom201110}%
  \BibitemOpen
  \bibfield  {author} {\bibinfo {author} {\bibfnamefont {J.~E.}\ \bibnamefont
  {Carlstrom}} \emph {et~al.},\ }\bibfield  {title} {\bibinfo {title} {{The 10
  Meter South Pole Telescope}},\ }\href {https://doi.org/10.1086/659879}
  {\bibfield  {journal} {\bibinfo  {journal} {Publ. Astron. Soc. Pac.}\
  }\textbf {\bibinfo {volume} {123}},\ \bibinfo {pages} {568} (\bibinfo {year}
  {2011})},\ \Eprint {https://arxiv.org/abs/0907.4445} {arXiv:0907.4445}
  \BibitemShut {NoStop}%
\bibitem [{\citenamefont {Pan}\ \emph {et~al.}(2019)\citenamefont {Pan},
  \citenamefont {Liu}, \citenamefont {Basu~Thakur}, \citenamefont {Benson},
  \citenamefont {Fixsen}, \citenamefont {Goksu}, \citenamefont {Rath},\ and\
  \citenamefont {Meyer}}]{Pan:2019omw}%
  \BibitemOpen
  \bibfield  {author} {\bibinfo {author} {\bibfnamefont {Z.}~\bibnamefont
  {Pan}}, \bibinfo {author} {\bibfnamefont {M.}~\bibnamefont {Liu}}, \bibinfo
  {author} {\bibfnamefont {R.}~\bibnamefont {Basu~Thakur}}, \bibinfo {author}
  {\bibfnamefont {B.~A.}\ \bibnamefont {Benson}}, \bibinfo {author}
  {\bibfnamefont {D.~J.}\ \bibnamefont {Fixsen}}, \bibinfo {author}
  {\bibfnamefont {H.}~\bibnamefont {Goksu}}, \bibinfo {author} {\bibfnamefont
  {E.}~\bibnamefont {Rath}},\ and\ \bibinfo {author} {\bibfnamefont {S.~S.}\
  \bibnamefont {Meyer}},\ }\bibfield  {title} {\bibinfo {title} {{Compact
  millimeter-wavelength Fourier-transform spectrometer}},\ }\href
  {https://doi.org/10.1364/AO.58.006257} {\bibfield  {journal} {\bibinfo
  {journal} {Appl. Opt.}\ }\textbf {\bibinfo {volume} {58}},\ \bibinfo {pages}
  {6257} (\bibinfo {year} {2019})},\ \Eprint {https://arxiv.org/abs/1905.07399}
  {arXiv:1905.07399 [astro-ph.IM]} \BibitemShut {NoStop}%
\bibitem [{\citenamefont {Thornton}\ \emph {et~al.}(2016)\citenamefont
  {Thornton}, \citenamefont {Ade}, \citenamefont {Aiola}, \citenamefont
  {Angile}, \citenamefont {Amiri}, \citenamefont {Beall}, \citenamefont
  {Becker}, \citenamefont {Cho}, \citenamefont {Choi}, \citenamefont {Corlies}
  \emph {et~al.}}]{thornton2016atacama}%
  \BibitemOpen
  \bibfield  {author} {\bibinfo {author} {\bibfnamefont {R.}~\bibnamefont
  {Thornton}}, \bibinfo {author} {\bibfnamefont {P.}~\bibnamefont {Ade}},
  \bibinfo {author} {\bibfnamefont {S.}~\bibnamefont {Aiola}}, \bibinfo
  {author} {\bibfnamefont {F.}~\bibnamefont {Angile}}, \bibinfo {author}
  {\bibfnamefont {M.}~\bibnamefont {Amiri}}, \bibinfo {author} {\bibfnamefont
  {J.}~\bibnamefont {Beall}}, \bibinfo {author} {\bibfnamefont
  {D.}~\bibnamefont {Becker}}, \bibinfo {author} {\bibfnamefont
  {H.}~\bibnamefont {Cho}}, \bibinfo {author} {\bibfnamefont {S.}~\bibnamefont
  {Choi}}, \bibinfo {author} {\bibfnamefont {P.}~\bibnamefont {Corlies}}, \emph
  {et~al.},\ }\bibfield  {title} {\bibinfo {title} {{The Atacama Cosmology
  Telescope: The polarization-sensitive ACTPol instrument}},\ }\href
  {https://doi.org/10.3847/1538-4365/227/2/21} {\bibfield  {journal} {\bibinfo
  {journal} {Astrophys. J., Suppl. Ser.}\ }\textbf {\bibinfo {volume} {227}},\
  \bibinfo {pages} {21} (\bibinfo {year} {2016})},\ \Eprint
  {https://arxiv.org/abs/1605.06569} {arXiv:1605.06569} \BibitemShut {NoStop}%
\bibitem [{\citenamefont {Matsuda}\ \emph {et~al.}(2019)\citenamefont {Matsuda}
  \emph {et~al.}}]{matsuda2019polarbear}%
  \BibitemOpen
  \bibfield  {author} {\bibinfo {author} {\bibfnamefont {F.}~\bibnamefont
  {Matsuda}} \emph {et~al.},\ }\bibfield  {title} {\bibinfo {title} {{The
  POLARBEAR Fourier transform spectrometer calibrator and spectroscopic
  characterization of the POLARBEAR instrument}},\ }\href
  {https://doi.org/10.1063/1.5095160} {\bibfield  {journal} {\bibinfo
  {journal} {Rev. Sci. Instrum.}\ }\textbf {\bibinfo {volume} {90}},\ \bibinfo
  {pages} {115115} (\bibinfo {year} {2019})},\ \Eprint
  {https://arxiv.org/abs/1904.02901} {arXiv:1904.02901} \BibitemShut {NoStop}%
\bibitem [{\citenamefont {Karkare}\ \emph {et~al.}(2014)\citenamefont {Karkare}
  \emph {et~al.}}]{BICEP2014}%
  \BibitemOpen
  \bibfield  {author} {\bibinfo {author} {\bibfnamefont {K.~S.}\ \bibnamefont
  {Karkare}} \emph {et~al.},\ }\bibfield  {title} {\bibinfo {title} {{Keck
  array and BICEP3: spectral characterization of 5000+ detectors}},\ }in\ \href
  {https://doi.org/10.1117/12.2056779} {\emph {\bibinfo {booktitle}
  {Millimeter, Submillimeter, and Far-Infrared Detectors and Instrumentation
  for Astronomy VII}}},\ Vol.\ \bibinfo {volume} {9153},\ \bibinfo {editor}
  {edited by\ \bibinfo {editor} {\bibfnamefont {W.~S.}\ \bibnamefont
  {Holland}}\ and\ \bibinfo {editor} {\bibfnamefont {J.}~\bibnamefont
  {Zmuidzinas}}},\ \bibinfo {organization} {International Society for Optics
  and Photonics}\ (\bibinfo  {publisher} {SPIE},\ \bibinfo {year} {2014})\ pp.\
  \bibinfo {pages} {1027 -- 1037}\BibitemShut {NoStop}%
\bibitem [{\citenamefont {Tonouchi}(2007)}]{tonouchi2007cutting}%
  \BibitemOpen
  \bibfield  {author} {\bibinfo {author} {\bibfnamefont {M.}~\bibnamefont
  {Tonouchi}},\ }\bibfield  {title} {\bibinfo {title} {Cutting-edge terahertz
  technology},\ }\href {https://doi.org/10.1038/nphoton.2007.3} {\bibfield
  {journal} {\bibinfo  {journal} {Nat. Photonics}\ }\textbf {\bibinfo {volume}
  {1}},\ \bibinfo {pages} {97} (\bibinfo {year} {2007})}\BibitemShut {NoStop}%
\bibitem [{\citenamefont {Zouaghi}\ \emph {et~al.}(2013)\citenamefont
  {Zouaghi}, \citenamefont {Thomson}, \citenamefont {Rabia}, \citenamefont
  {Hahn}, \citenamefont {Blank},\ and\ \citenamefont
  {Roskos}}]{zouaghi2013broadband}%
  \BibitemOpen
  \bibfield  {author} {\bibinfo {author} {\bibfnamefont {W.}~\bibnamefont
  {Zouaghi}}, \bibinfo {author} {\bibfnamefont {M.}~\bibnamefont {Thomson}},
  \bibinfo {author} {\bibfnamefont {K.}~\bibnamefont {Rabia}}, \bibinfo
  {author} {\bibfnamefont {R.}~\bibnamefont {Hahn}}, \bibinfo {author}
  {\bibfnamefont {V.}~\bibnamefont {Blank}},\ and\ \bibinfo {author}
  {\bibfnamefont {H.}~\bibnamefont {Roskos}},\ }\bibfield  {title} {\bibinfo
  {title} {Broadband terahertz spectroscopy: principles, fundamental research
  and potential for industrial applications},\ }\href
  {https://doi.org/10.1088/0143-0807/34/6/S179} {\bibfield  {journal} {\bibinfo
   {journal} {Eur. J. Phys.}\ }\textbf {\bibinfo {volume} {34}},\ \bibinfo
  {pages} {S179} (\bibinfo {year} {2013})}\BibitemShut {NoStop}%
\bibitem [{\citenamefont {Dhillon}\ \emph {et~al.}(2017)\citenamefont {Dhillon}
  \emph {et~al.}}]{dhillon20172017}%
  \BibitemOpen
  \bibfield  {author} {\bibinfo {author} {\bibfnamefont {S.~S.}\ \bibnamefont
  {Dhillon}} \emph {et~al.},\ }\bibfield  {title} {\bibinfo {title} {The 2017
  terahertz science and technology roadmap},\ }\href
  {https://doi.org/10.1088/1361-6463/50/4/043001} {\bibfield  {journal}
  {\bibinfo  {journal} {J. Phys. D: Appl. Phys}\ }\textbf {\bibinfo {volume}
  {50}},\ \bibinfo {pages} {043001} (\bibinfo {year} {2017})}\BibitemShut
  {NoStop}%
\bibitem [{\citenamefont {Carelli}\ \emph {et~al.}(2017)\citenamefont
  {Carelli}, \citenamefont {Chiarello}, \citenamefont {Torrioli},\ and\
  \citenamefont {Castellano}}]{Carelli2017}%
  \BibitemOpen
  \bibfield  {author} {\bibinfo {author} {\bibfnamefont {P.}~\bibnamefont
  {Carelli}}, \bibinfo {author} {\bibfnamefont {F.}~\bibnamefont {Chiarello}},
  \bibinfo {author} {\bibfnamefont {G.}~\bibnamefont {Torrioli}},\ and\
  \bibinfo {author} {\bibfnamefont {M.~G.}\ \bibnamefont {Castellano}},\
  }\bibfield  {title} {\bibinfo {title} {{THz Discrimination of Materials:
  Development of an Apparatus Based on Room Temperature Detection and
  Metasurfaces Selective Filters}},\ }\href
  {https://doi.org/10.1007/s10762-016-0343-0} {\bibfield  {journal} {\bibinfo
  {journal} {J. Infrared Millim. Terahertz Waves}\ }\textbf {\bibinfo {volume}
  {38}},\ \bibinfo {pages} {303} (\bibinfo {year} {2017})},\ \Eprint
  {https://arxiv.org/abs/1607.04512} {arXiv:1607.04512} \BibitemShut {NoStop}%
\bibitem [{\citenamefont {Martini}\ \emph {et~al.}(2020)\citenamefont
  {Martini}, \citenamefont {Giovine}, \citenamefont {Chiarello},\ and\
  \citenamefont {Carelli}}]{Martini:2020owl}%
  \BibitemOpen
  \bibfield  {author} {\bibinfo {author} {\bibfnamefont {F.}~\bibnamefont
  {Martini}}, \bibinfo {author} {\bibfnamefont {E.}~\bibnamefont {Giovine}},
  \bibinfo {author} {\bibfnamefont {F.}~\bibnamefont {Chiarello}},\ and\
  \bibinfo {author} {\bibfnamefont {P.}~\bibnamefont {Carelli}},\ }\bibfield
  {title} {\bibinfo {title} {{A THz spectrometer using band pass filters}},\
  }\href {https://doi.org/10.3390/instruments4030024} {\bibfield  {journal}
  {\bibinfo  {journal} {Instruments}\ }\textbf {\bibinfo {volume} {4}},\
  \bibinfo {pages} {24} (\bibinfo {year} {2020})},\ \Eprint
  {https://arxiv.org/abs/2005.10540} {arXiv:2005.10540 [physics.ins-det]}
  \BibitemShut {NoStop}%
\bibitem [{\citenamefont {Pickwell}\ and\ \citenamefont
  {Wallace}(2006)}]{pickwell2006biomedical}%
  \BibitemOpen
  \bibfield  {author} {\bibinfo {author} {\bibfnamefont {E.}~\bibnamefont
  {Pickwell}}\ and\ \bibinfo {author} {\bibfnamefont {V.}~\bibnamefont
  {Wallace}},\ }\bibfield  {title} {\bibinfo {title} {Biomedical applications
  of terahertz technology},\ }\href
  {https://doi.org/10.1088/0022-3727/39/17/R01} {\bibfield  {journal} {\bibinfo
   {journal} {J. Phys. D: Appl. Phys}\ }\textbf {\bibinfo {volume} {39}},\
  \bibinfo {pages} {R301} (\bibinfo {year} {2006})}\BibitemShut {NoStop}%
\bibitem [{\citenamefont {Reid}\ \emph {et~al.}(2010)\citenamefont {Reid},
  \citenamefont {Pickwell-MacPherson}, \citenamefont {Laufer}, \citenamefont
  {Gibson}, \citenamefont {Hebden},\ and\ \citenamefont
  {Wallace}}]{reid2010accuracy}%
  \BibitemOpen
  \bibfield  {author} {\bibinfo {author} {\bibfnamefont {C.~B.}\ \bibnamefont
  {Reid}}, \bibinfo {author} {\bibfnamefont {E.}~\bibnamefont
  {Pickwell-MacPherson}}, \bibinfo {author} {\bibfnamefont {J.~G.}\
  \bibnamefont {Laufer}}, \bibinfo {author} {\bibfnamefont {A.~P.}\
  \bibnamefont {Gibson}}, \bibinfo {author} {\bibfnamefont {J.~C.}\
  \bibnamefont {Hebden}},\ and\ \bibinfo {author} {\bibfnamefont {V.~P.}\
  \bibnamefont {Wallace}},\ }\bibfield  {title} {\bibinfo {title} {{Accuracy
  and resolution of THz reflection spectroscopy for medical imaging}},\ }\href
  {https://doi.org/doi.org/10.1088/0031-9155/55/16/013} {\bibfield  {journal}
  {\bibinfo  {journal} {Phys. Med. Biol.}\ }\textbf {\bibinfo {volume} {55}},\
  \bibinfo {pages} {4825} (\bibinfo {year} {2010})}\BibitemShut {NoStop}%
\bibitem [{\citenamefont {Shen}(2011)}]{shen2011terahertz}%
  \BibitemOpen
  \bibfield  {author} {\bibinfo {author} {\bibfnamefont {Y.-C.}\ \bibnamefont
  {Shen}},\ }\bibfield  {title} {\bibinfo {title} {{Terahertz pulsed
  spectroscopy and imaging for pharmaceutical applications: A review}},\ }\href
  {https://doi.org/10.1016/j.ijpharm.2011.01.012} {\bibfield  {journal}
  {\bibinfo  {journal} {Int. J. Pharm.}\ }\textbf {\bibinfo {volume} {417}},\
  \bibinfo {pages} {48} (\bibinfo {year} {2011})}\BibitemShut {NoStop}%
\bibitem [{\citenamefont {Taylor}\ \emph {et~al.}(2011)\citenamefont {Taylor}
  \emph {et~al.}}]{Taylor2011}%
  \BibitemOpen
  \bibfield  {author} {\bibinfo {author} {\bibfnamefont {Z.~D.}\ \bibnamefont
  {Taylor}} \emph {et~al.},\ }\bibfield  {title} {\bibinfo {title} {{THz
  Medical Imaging: in vivo Hydration Sensing}},\ }\href
  {https://doi.org/10.1109/TTHZ.2011.2159551} {\bibfield  {journal} {\bibinfo
  {journal} {IEEE Trans. Terahertz. Sci. Technol.}\ }\textbf {\bibinfo {volume}
  {1}},\ \bibinfo {pages} {201} (\bibinfo {year} {2011})}\BibitemShut {NoStop}%
\bibitem [{\citenamefont {Hintzsche}\ and\ \citenamefont
  {Stopper}(2012)}]{Hintzsche2012}%
  \BibitemOpen
  \bibfield  {author} {\bibinfo {author} {\bibfnamefont {H.}~\bibnamefont
  {Hintzsche}}\ and\ \bibinfo {author} {\bibfnamefont {H.}~\bibnamefont
  {Stopper}},\ }\bibfield  {title} {\bibinfo {title} {{Effects of Terahertz
  Radiation on Biological Systems}},\ }\href
  {https://doi.org/10.1080/10643389.2011.574206} {\bibfield  {journal}
  {\bibinfo  {journal} {Crit. Rev. Environ. Sci. Technol.}\ }\textbf {\bibinfo
  {volume} {42}},\ \bibinfo {pages} {2408} (\bibinfo {year}
  {2012})}\BibitemShut {NoStop}%
\bibitem [{\citenamefont {Williams}\ \emph {et~al.}(2013)\citenamefont
  {Williams}, \citenamefont {Aschaffenburg}, \citenamefont {Ofori-Okai},\ and\
  \citenamefont {Schmuttenmaer}}]{williams2013intermolecular}%
  \BibitemOpen
  \bibfield  {author} {\bibinfo {author} {\bibfnamefont {M.~R.}\ \bibnamefont
  {Williams}}, \bibinfo {author} {\bibfnamefont {D.~J.}\ \bibnamefont
  {Aschaffenburg}}, \bibinfo {author} {\bibfnamefont {B.~K.}\ \bibnamefont
  {Ofori-Okai}},\ and\ \bibinfo {author} {\bibfnamefont {C.~A.}\ \bibnamefont
  {Schmuttenmaer}},\ }\bibfield  {title} {\bibinfo {title} {{Intermolecular
  vibrations in hydrophobic amino acid crystals: experiments and
  calculations}},\ }\href {https://doi.org/10.1021/jp406730a} {\bibfield
  {journal} {\bibinfo  {journal} {J. Phys. Chem. B}\ }\textbf {\bibinfo
  {volume} {117}},\ \bibinfo {pages} {10444} (\bibinfo {year}
  {2013})}\BibitemShut {NoStop}%
\bibitem [{\citenamefont {Hishida}\ and\ \citenamefont
  {Tanaka}(2011)}]{PhysRevLett.106.158102}%
  \BibitemOpen
  \bibfield  {author} {\bibinfo {author} {\bibfnamefont {M.}~\bibnamefont
  {Hishida}}\ and\ \bibinfo {author} {\bibfnamefont {K.}~\bibnamefont
  {Tanaka}},\ }\bibfield  {title} {\bibinfo {title} {{Long-Range Hydration
  Effect of Lipid Membrane Studied by Terahertz Time-Domain Spectroscopy}},\
  }\href {https://doi.org/10.1103/PhysRevLett.106.158102} {\bibfield  {journal}
  {\bibinfo  {journal} {Phys. Rev. Lett.}\ }\textbf {\bibinfo {volume} {106}},\
  \bibinfo {pages} {158102} (\bibinfo {year} {2011})}\BibitemShut {NoStop}%
\bibitem [{\citenamefont {Hindle}\ \emph {et~al.}(2008)\citenamefont {Hindle},
  \citenamefont {Cuisset}, \citenamefont {Bocquet},\ and\ \citenamefont
  {Mouret}}]{hindle2008continuous}%
  \BibitemOpen
  \bibfield  {author} {\bibinfo {author} {\bibfnamefont {F.}~\bibnamefont
  {Hindle}}, \bibinfo {author} {\bibfnamefont {A.}~\bibnamefont {Cuisset}},
  \bibinfo {author} {\bibfnamefont {R.}~\bibnamefont {Bocquet}},\ and\ \bibinfo
  {author} {\bibfnamefont {G.}~\bibnamefont {Mouret}},\ }\bibfield  {title}
  {\bibinfo {title} {Continuous-wave terahertz by photomixing: applications to
  gas phase pollutant detection and quantification},\ }\href
  {https://doi.org/10.1016/j.crhy.2007.07.009} {\bibfield  {journal} {\bibinfo
  {journal} {C. R. Phys.}\ }\textbf {\bibinfo {volume} {9}},\ \bibinfo {pages}
  {262} (\bibinfo {year} {2008})}\BibitemShut {NoStop}%
\bibitem [{\citenamefont {Slocum}\ \emph {et~al.}(2013)\citenamefont {Slocum},
  \citenamefont {Slingerland}, \citenamefont {Giles},\ and\ \citenamefont
  {Goyette}}]{slocum2013atmospheric}%
  \BibitemOpen
  \bibfield  {author} {\bibinfo {author} {\bibfnamefont {D.~M.}\ \bibnamefont
  {Slocum}}, \bibinfo {author} {\bibfnamefont {E.~J.}\ \bibnamefont
  {Slingerland}}, \bibinfo {author} {\bibfnamefont {R.~H.}\ \bibnamefont
  {Giles}},\ and\ \bibinfo {author} {\bibfnamefont {T.~M.}\ \bibnamefont
  {Goyette}},\ }\bibfield  {title} {\bibinfo {title} {Atmospheric absorption of
  terahertz radiation and water vapor continuum effects},\ }\href
  {https://doi.org/10.1016/j.jqsrt.2013.04.022} {\bibfield  {journal} {\bibinfo
   {journal} {J Quant. Spectrosc. Radiat. Transf.}\ }\textbf {\bibinfo {volume}
  {127}},\ \bibinfo {pages} {49} (\bibinfo {year} {2013})}\BibitemShut
  {NoStop}%
\bibitem [{\citenamefont {Hsieh}\ \emph {et~al.}(2016)\citenamefont {Hsieh}
  \emph {et~al.}}]{hsieh2016dynamic}%
  \BibitemOpen
  \bibfield  {author} {\bibinfo {author} {\bibfnamefont {Y.-D.}\ \bibnamefont
  {Hsieh}} \emph {et~al.},\ }\bibfield  {title} {\bibinfo {title} {{Dynamic
  terahertz spectroscopy of gas molecules mixed with unwanted aerosol under
  atmospheric pressure using fibre-based asynchronous-optical-sampling
  terahertz time-domain spectroscopy}},\ }\href
  {https://doi.org/10.1038/srep28114} {\bibfield  {journal} {\bibinfo
  {journal} {Sci. Rep.}\ }\textbf {\bibinfo {volume} {6}},\ \bibinfo {pages}
  {1} (\bibinfo {year} {2016})}\BibitemShut {NoStop}%
\bibitem [{\citenamefont {Chan}\ \emph {et~al.}(2007)\citenamefont {Chan},
  \citenamefont {Deibel},\ and\ \citenamefont {Mittleman}}]{chan2007imaging}%
  \BibitemOpen
  \bibfield  {author} {\bibinfo {author} {\bibfnamefont {W.~L.}\ \bibnamefont
  {Chan}}, \bibinfo {author} {\bibfnamefont {J.}~\bibnamefont {Deibel}},\ and\
  \bibinfo {author} {\bibfnamefont {D.~M.}\ \bibnamefont {Mittleman}},\
  }\bibfield  {title} {\bibinfo {title} {Imaging with terahertz radiation},\
  }\href {https://doi.org/10.1088/0034-4885/70/8/R02} {\bibfield  {journal}
  {\bibinfo  {journal} {Rep. Prog. Phys.}\ }\textbf {\bibinfo {volume} {70}},\
  \bibinfo {pages} {1325} (\bibinfo {year} {2007})}\BibitemShut {NoStop}%
\bibitem [{\citenamefont {{Kemp}}(2011)}]{Kemp2011}%
  \BibitemOpen
  \bibfield  {author} {\bibinfo {author} {\bibfnamefont {M.~C.}\ \bibnamefont
  {{Kemp}}},\ }\bibfield  {title} {\bibinfo {title} {Explosives detection by
  terahertz spectroscopy--a bridge too far?},\ }\href
  {https://doi.org/10.1109/TTHZ.2011.2159647} {\bibfield  {journal} {\bibinfo
  {journal} {IEEE Trans. Terahertz. Sci. Technol.}\ }\textbf {\bibinfo {volume}
  {1}},\ \bibinfo {pages} {282} (\bibinfo {year} {2011})}\BibitemShut {NoStop}%
\bibitem [{\citenamefont {{Cooper}}\ and\ \citenamefont
  {{Chattopadhyay}}(2014)}]{Cooper2014}%
  \BibitemOpen
  \bibfield  {author} {\bibinfo {author} {\bibfnamefont {K.~B.}\ \bibnamefont
  {{Cooper}}}\ and\ \bibinfo {author} {\bibfnamefont {G.}~\bibnamefont
  {{Chattopadhyay}}},\ }\bibfield  {title} {\bibinfo {title}
  {{Submillimeter-Wave Radar: Solid-State System Design and Applications}},\
  }\href {https://doi.org/10.1109/MMM.2014.2356092} {\bibfield  {journal}
  {\bibinfo  {journal} {IEEE Microw. Mag.}\ }\textbf {\bibinfo {volume} {15}},\
  \bibinfo {pages} {51} (\bibinfo {year} {2014})}\BibitemShut {NoStop}%
\bibitem [{\citenamefont {Heinz}\ \emph {et~al.}(2015)\citenamefont {Heinz},
  \citenamefont {May}, \citenamefont {Born}, \citenamefont {Zieger},
  \citenamefont {Anders}, \citenamefont {Zakosarenko}, \citenamefont {Meyer},\
  and\ \citenamefont {Sch{\"a}ffel}}]{heinz2015passive}%
  \BibitemOpen
  \bibfield  {author} {\bibinfo {author} {\bibfnamefont {E.}~\bibnamefont
  {Heinz}}, \bibinfo {author} {\bibfnamefont {T.}~\bibnamefont {May}}, \bibinfo
  {author} {\bibfnamefont {D.}~\bibnamefont {Born}}, \bibinfo {author}
  {\bibfnamefont {G.}~\bibnamefont {Zieger}}, \bibinfo {author} {\bibfnamefont
  {S.}~\bibnamefont {Anders}}, \bibinfo {author} {\bibfnamefont
  {V.}~\bibnamefont {Zakosarenko}}, \bibinfo {author} {\bibfnamefont {H.-G.}\
  \bibnamefont {Meyer}},\ and\ \bibinfo {author} {\bibfnamefont
  {C.}~\bibnamefont {Sch{\"a}ffel}},\ }\bibfield  {title} {\bibinfo {title}
  {{Passive 350 GHz video imaging systems for security applications}},\ }\href
  {https://doi.org/10.1007/s10762-015-0170-8} {\bibfield  {journal} {\bibinfo
  {journal} {J. Infrared Millim. Terahertz Waves}\ }\textbf {\bibinfo {volume}
  {36}},\ \bibinfo {pages} {879} (\bibinfo {year} {2015})}\BibitemShut
  {NoStop}%
\bibitem [{\citenamefont {Federici}\ and\ \citenamefont
  {Moeller}(2010)}]{federici2010review}%
  \BibitemOpen
  \bibfield  {author} {\bibinfo {author} {\bibfnamefont {J.}~\bibnamefont
  {Federici}}\ and\ \bibinfo {author} {\bibfnamefont {L.}~\bibnamefont
  {Moeller}},\ }\bibfield  {title} {\bibinfo {title} {Review of terahertz and
  subterahertz wireless communications},\ }\href
  {https://doi.org/10.1063/1.3386413} {\bibfield  {journal} {\bibinfo
  {journal} {J. Appl. Phys.}\ }\textbf {\bibinfo {volume} {107}},\ \bibinfo
  {pages} {6} (\bibinfo {year} {2010})}\BibitemShut {NoStop}%
\bibitem [{\citenamefont {Akyildiz}\ \emph {et~al.}(2014)\citenamefont
  {Akyildiz}, \citenamefont {Jornet},\ and\ \citenamefont
  {Han}}]{akyildiz2014terahertz}%
  \BibitemOpen
  \bibfield  {author} {\bibinfo {author} {\bibfnamefont {I.~F.}\ \bibnamefont
  {Akyildiz}}, \bibinfo {author} {\bibfnamefont {J.~M.}\ \bibnamefont
  {Jornet}},\ and\ \bibinfo {author} {\bibfnamefont {C.}~\bibnamefont {Han}},\
  }\bibfield  {title} {\bibinfo {title} {{Terahertz band: Next frontier for
  wireless communications}},\ }\href
  {https://doi.org/10.1016/j.phycom.2014.01.006} {\bibfield  {journal}
  {\bibinfo  {journal} {Phys. Commun.}\ }\textbf {\bibinfo {volume} {12}},\
  \bibinfo {pages} {16} (\bibinfo {year} {2014})}\BibitemShut {NoStop}%
\bibitem [{\citenamefont {{Seeds}}\ \emph {et~al.}(2015)\citenamefont
  {{Seeds}}, \citenamefont {{Shams}}, \citenamefont {{Fice}},\ and\
  \citenamefont {{Renaud}}}]{Seeds2015}%
  \BibitemOpen
  \bibfield  {author} {\bibinfo {author} {\bibfnamefont {A.~J.}\ \bibnamefont
  {{Seeds}}}, \bibinfo {author} {\bibfnamefont {H.}~\bibnamefont {{Shams}}},
  \bibinfo {author} {\bibfnamefont {M.~J.}\ \bibnamefont {{Fice}}},\ and\
  \bibinfo {author} {\bibfnamefont {C.~C.}\ \bibnamefont {{Renaud}}},\
  }\bibfield  {title} {\bibinfo {title} {{TeraHertz Photonics for Wireless
  Communications}},\ }\href {https://doi.org/10.1109/JLT.2014.2355137}
  {\bibfield  {journal} {\bibinfo  {journal} {J. Light. Technol.}\ }\textbf
  {\bibinfo {volume} {33}},\ \bibinfo {pages} {579} (\bibinfo {year}
  {2015})}\BibitemShut {NoStop}%
\bibitem [{\citenamefont {Jaeckel}\ and\ \citenamefont
  {Ringwald}(2010)}]{Jaeckel:2010ni}%
  \BibitemOpen
  \bibfield  {author} {\bibinfo {author} {\bibfnamefont {J.}~\bibnamefont
  {Jaeckel}}\ and\ \bibinfo {author} {\bibfnamefont {A.}~\bibnamefont
  {Ringwald}},\ }\bibfield  {title} {\bibinfo {title} {{The Low-Energy Frontier
  of Particle Physics}},\ }\href
  {https://doi.org/10.1146/annurev.nucl.012809.104433} {\bibfield  {journal}
  {\bibinfo  {journal} {Ann. Rev. Nucl. Part. Sci.}\ }\textbf {\bibinfo
  {volume} {60}},\ \bibinfo {pages} {405} (\bibinfo {year} {2010})},\ \Eprint
  {https://arxiv.org/abs/1002.0329} {arXiv:1002.0329 [hep-ph]} \BibitemShut
  {NoStop}%
\bibitem [{\citenamefont {Essig}\ \emph {et~al.}(2013)\citenamefont {Essig}
  \emph {et~al.}}]{Essig:2013lka}%
  \BibitemOpen
  \bibfield  {author} {\bibinfo {author} {\bibfnamefont {R.}~\bibnamefont
  {Essig}} \emph {et~al.},\ }\bibfield  {title} {\bibinfo {title} {{Working
  Group Report: New Light Weakly Coupled Particles}},\ }in\ \href
  {https://inspirehep.net/record/1263039/files/arXiv:1311.0029.pdf} {\emph
  {\bibinfo {booktitle} {{Community Summer Study 2013: Snowmass on the
  Mississippi}}}}\ (\bibinfo {year} {2013})\ \Eprint
  {https://arxiv.org/abs/1311.0029} {arXiv:1311.0029 [hep-ph]} \BibitemShut
  {NoStop}%
\bibitem [{\citenamefont {Baker}\ \emph {et~al.}(2013)\citenamefont {Baker}
  \emph {et~al.}}]{Baker:2013zta}%
  \BibitemOpen
  \bibfield  {author} {\bibinfo {author} {\bibfnamefont {K.}~\bibnamefont
  {Baker}} \emph {et~al.},\ }\bibfield  {title} {\bibinfo {title} {{The quest
  for axions and other new light particles}},\ }\href
  {https://doi.org/10.1002/andp.201300727} {\bibfield  {journal} {\bibinfo
  {journal} {Annalen Phys.}\ }\textbf {\bibinfo {volume} {525}},\ \bibinfo
  {pages} {A93} (\bibinfo {year} {2013})},\ \Eprint
  {https://arxiv.org/abs/1306.2841} {arXiv:1306.2841 [hep-ph]} \BibitemShut
  {NoStop}%
\bibitem [{\citenamefont {Battaglieri}\ \emph {et~al.}(2017)\citenamefont
  {Battaglieri} \emph {et~al.}}]{Battaglieri:2017aum}%
  \BibitemOpen
  \bibfield  {author} {\bibinfo {author} {\bibfnamefont {M.}~\bibnamefont
  {Battaglieri}} \emph {et~al.},\ }\bibfield  {title} {\bibinfo {title} {{New
  Ideas in Dark Matter 2017: Community Report}},\ }in\ \href
  {http://lss.fnal.gov/archive/2017/conf/fermilab-conf-17-282-ae-ppd-t.pdf}
  {\emph {\bibinfo {booktitle} {{U.S. Cosmic Visions}}}}\ (\bibinfo {year}
  {2017})\ \Eprint {https://arxiv.org/abs/1707.04591} {arXiv:1707.04591
  [hep-ph]} \BibitemShut {NoStop}%
\bibitem [{\citenamefont {Irastorza}\ and\ \citenamefont
  {Redondo}(2018)}]{Irastorza:2018dyq}%
  \BibitemOpen
  \bibfield  {author} {\bibinfo {author} {\bibfnamefont {I.~G.}\ \bibnamefont
  {Irastorza}}\ and\ \bibinfo {author} {\bibfnamefont {J.}~\bibnamefont
  {Redondo}},\ }\bibfield  {title} {\bibinfo {title} {{New experimental
  approaches in the search for axion-like particles}},\ }\href
  {https://doi.org/10.1016/j.ppnp.2018.05.003} {\bibfield  {journal} {\bibinfo
  {journal} {Prog. Part. Nucl. Phys.}\ }\textbf {\bibinfo {volume} {102}},\
  \bibinfo {pages} {89} (\bibinfo {year} {2018})},\ \Eprint
  {https://arxiv.org/abs/1801.08127} {arXiv:1801.08127 [hep-ph]} \BibitemShut
  {NoStop}%
\bibitem [{\citenamefont {Sikivie}(1983)}]{Sikivie:1983ip}%
  \BibitemOpen
  \bibfield  {author} {\bibinfo {author} {\bibfnamefont {P.}~\bibnamefont
  {Sikivie}},\ }\bibfield  {title} {\bibinfo {title} {{Experimental Tests of
  the Invisible Axion}},\ }\href {https://doi.org/10.1103/PhysRevLett.51.1415}
  {\bibfield  {journal} {\bibinfo  {journal} {Phys. Rev. Lett.}\ }\textbf
  {\bibinfo {volume} {51}},\ \bibinfo {pages} {1415} (\bibinfo {year}
  {1983})},\ \bibinfo {note} {[Erratum: Phys.Rev.Lett. 52, 695
  (1984)]}\BibitemShut {NoStop}%
\bibitem [{\citenamefont {De~Panfilis}\ \emph {et~al.}(1987)\citenamefont
  {De~Panfilis} \emph {et~al.}}]{DePanfilis:1987dk}%
  \BibitemOpen
  \bibfield  {author} {\bibinfo {author} {\bibfnamefont {S.}~\bibnamefont
  {De~Panfilis}} \emph {et~al.},\ }\bibfield  {title} {\bibinfo {title}
  {{Limits on the Abundance and Coupling of Cosmic Axions at $4.5< m_a <
  5.0~\mu$eV}},\ }\href {https://doi.org/10.1103/PhysRevLett.59.839} {\bibfield
   {journal} {\bibinfo  {journal} {Phys. Rev. Lett.}\ }\textbf {\bibinfo
  {volume} {59}},\ \bibinfo {pages} {839} (\bibinfo {year} {1987})}\BibitemShut
  {NoStop}%
\bibitem [{\citenamefont {Wuensch}\ \emph {et~al.}(1989)\citenamefont {Wuensch}
  \emph {et~al.}}]{Wuensch:1989sa}%
  \BibitemOpen
  \bibfield  {author} {\bibinfo {author} {\bibfnamefont {W.}~\bibnamefont
  {Wuensch}} \emph {et~al.},\ }\bibfield  {title} {\bibinfo {title} {{Results
  of a Laboratory Search for Cosmic Axions and Other Weakly Coupled Light
  Particles}},\ }\href {https://doi.org/10.1103/PhysRevD.40.3153} {\bibfield
  {journal} {\bibinfo  {journal} {Phys. Rev. D}\ }\textbf {\bibinfo {volume}
  {40}},\ \bibinfo {pages} {3153} (\bibinfo {year} {1989})}\BibitemShut
  {NoStop}%
\bibitem [{\citenamefont {Hagmann}\ \emph {et~al.}(1990)\citenamefont
  {Hagmann}, \citenamefont {Sikivie}, \citenamefont {Sullivan},\ and\
  \citenamefont {Tanner}}]{Hagmann:1990tj}%
  \BibitemOpen
  \bibfield  {author} {\bibinfo {author} {\bibfnamefont {C.}~\bibnamefont
  {Hagmann}}, \bibinfo {author} {\bibfnamefont {P.}~\bibnamefont {Sikivie}},
  \bibinfo {author} {\bibfnamefont {N.}~\bibnamefont {Sullivan}},\ and\
  \bibinfo {author} {\bibfnamefont {D.}~\bibnamefont {Tanner}},\ }\bibfield
  {title} {\bibinfo {title} {{Results from a search for cosmic axions}},\
  }\href {https://doi.org/10.1103/PhysRevD.42.1297} {\bibfield  {journal}
  {\bibinfo  {journal} {Phys. Rev. D}\ }\textbf {\bibinfo {volume} {42}},\
  \bibinfo {pages} {1297} (\bibinfo {year} {1990})}\BibitemShut {NoStop}%
\bibitem [{\citenamefont {Asztalos}\ \emph {et~al.}(2001)\citenamefont
  {Asztalos} \emph {et~al.}}]{Asztalos:2001tf}%
  \BibitemOpen
  \bibfield  {author} {\bibinfo {author} {\bibfnamefont {S.~J.}\ \bibnamefont
  {Asztalos}} \emph {et~al.} (\bibinfo {collaboration} {ADMX}),\ }\bibfield
  {title} {\bibinfo {title} {{Large scale microwave cavity search for dark
  matter axions}},\ }\href {https://doi.org/10.1103/PhysRevD.64.092003}
  {\bibfield  {journal} {\bibinfo  {journal} {Phys.\ Rev.\ D}\ }\textbf
  {\bibinfo {volume} {64}},\ \bibinfo {pages} {092003} (\bibinfo {year}
  {2001})}\BibitemShut {NoStop}%
\bibitem [{\citenamefont {Asztalos}\ \emph {et~al.}(2002)\citenamefont
  {Asztalos} \emph {et~al.}}]{Asztalos:2001jk}%
  \BibitemOpen
  \bibfield  {author} {\bibinfo {author} {\bibfnamefont {S.~J.}\ \bibnamefont
  {Asztalos}} \emph {et~al.} (\bibinfo {collaboration} {ADMX}),\ }\bibfield
  {title} {\bibinfo {title} {{Experimental constraints on the axion dark matter
  halo density}},\ }\href {https://doi.org/10.1086/341130} {\bibfield
  {journal} {\bibinfo  {journal} {Astrophys. J. Lett.}\ }\textbf {\bibinfo
  {volume} {571}},\ \bibinfo {pages} {L27} (\bibinfo {year} {2002})},\ \Eprint
  {https://arxiv.org/abs/astro-ph/0104200} {arXiv:astro-ph/0104200}
  \BibitemShut {NoStop}%
\bibitem [{\citenamefont {Asztalos}\ \emph {et~al.}(2010)\citenamefont
  {Asztalos} \emph {et~al.}}]{Asztalos:2009yp}%
  \BibitemOpen
  \bibfield  {author} {\bibinfo {author} {\bibfnamefont {S.}~\bibnamefont
  {Asztalos}} \emph {et~al.} (\bibinfo {collaboration} {ADMX}),\ }\bibfield
  {title} {\bibinfo {title} {{A SQUID-based microwave cavity search for
  dark-matter axions}},\ }\href
  {https://doi.org/10.1103/PhysRevLett.104.041301} {\bibfield  {journal}
  {\bibinfo  {journal} {Phys.\ Rev.\ Lett.}\ }\textbf {\bibinfo {volume}
  {104}},\ \bibinfo {pages} {041301} (\bibinfo {year} {2010})},\ \Eprint
  {https://arxiv.org/abs/0910.5914} {arXiv:0910.5914 [astro-ph.CO]}
  \BibitemShut {NoStop}%
\bibitem [{\citenamefont {Wagner}\ \emph {et~al.}(2010)\citenamefont {Wagner}
  \emph {et~al.}}]{Wagner:2010mi}%
  \BibitemOpen
  \bibfield  {author} {\bibinfo {author} {\bibfnamefont {A.}~\bibnamefont
  {Wagner}} \emph {et~al.} (\bibinfo {collaboration} {ADMX}),\ }\bibfield
  {title} {\bibinfo {title} {{A Search for Hidden Sector Photons with ADMX}},\
  }\href {https://doi.org/10.1103/PhysRevLett.105.171801} {\bibfield  {journal}
  {\bibinfo  {journal} {Phys. Rev. Lett.}\ }\textbf {\bibinfo {volume} {105}},\
  \bibinfo {pages} {171801} (\bibinfo {year} {2010})},\ \Eprint
  {https://arxiv.org/abs/1007.3766} {arXiv:1007.3766 [hep-ex]} \BibitemShut
  {NoStop}%
\bibitem [{\citenamefont {Du}\ \emph {et~al.}(2018)\citenamefont {Du} \emph
  {et~al.}}]{Du:2018uak}%
  \BibitemOpen
  \bibfield  {author} {\bibinfo {author} {\bibfnamefont {N.}~\bibnamefont {Du}}
  \emph {et~al.} (\bibinfo {collaboration} {ADMX}),\ }\bibfield  {title}
  {\bibinfo {title} {{A Search for Invisible Axion Dark Matter with the Axion
  Dark Matter Experiment}},\ }\href
  {https://doi.org/10.1103/PhysRevLett.120.151301} {\bibfield  {journal}
  {\bibinfo  {journal} {Phys. Rev. Lett.}\ }\textbf {\bibinfo {volume} {120}},\
  \bibinfo {pages} {151301} (\bibinfo {year} {2018})},\ \Eprint
  {https://arxiv.org/abs/1804.05750} {arXiv:1804.05750 [hep-ex]} \BibitemShut
  {NoStop}%
\bibitem [{\citenamefont {Al~Kenany}\ \emph {et~al.}(2017)\citenamefont
  {Al~Kenany} \emph {et~al.}}]{Kenany:2016tta}%
  \BibitemOpen
  \bibfield  {author} {\bibinfo {author} {\bibfnamefont {S.}~\bibnamefont
  {Al~Kenany}} \emph {et~al.},\ }\bibfield  {title} {\bibinfo {title} {{Design
  and operational experience of a microwave cavity axion detector for the
  20--100 $\mu$eV range}},\ }\href {https://doi.org/10.1016/j.nima.2017.02.012}
  {\bibfield  {journal} {\bibinfo  {journal} {Nucl. Instrum. Meth. A}\ }\textbf
  {\bibinfo {volume} {854}},\ \bibinfo {pages} {11} (\bibinfo {year} {2017})},\
  \Eprint {https://arxiv.org/abs/1611.07123} {arXiv:1611.07123
  [physics.ins-det]} \BibitemShut {NoStop}%
\bibitem [{\citenamefont {Brubaker}\ \emph {et~al.}(2017)\citenamefont
  {Brubaker} \emph {et~al.}}]{Brubaker:2016ktl}%
  \BibitemOpen
  \bibfield  {author} {\bibinfo {author} {\bibfnamefont {B.}~\bibnamefont
  {Brubaker}} \emph {et~al.},\ }\bibfield  {title} {\bibinfo {title} {{First
  results from a microwave cavity axion search at 24 $\mu$eV}},\ }\href
  {https://doi.org/10.1103/PhysRevLett.118.061302} {\bibfield  {journal}
  {\bibinfo  {journal} {Phys. Rev. Lett.}\ }\textbf {\bibinfo {volume} {118}},\
  \bibinfo {pages} {061302} (\bibinfo {year} {2017})},\ \Eprint
  {https://arxiv.org/abs/1610.02580} {arXiv:1610.02580 [astro-ph.CO]}
  \BibitemShut {NoStop}%
\bibitem [{\citenamefont {Zhong}\ \emph {et~al.}(2018)\citenamefont {Zhong}
  \emph {et~al.}}]{Zhong:2018rsr}%
  \BibitemOpen
  \bibfield  {author} {\bibinfo {author} {\bibfnamefont {L.}~\bibnamefont
  {Zhong}} \emph {et~al.} (\bibinfo {collaboration} {HAYSTAC}),\ }\bibfield
  {title} {\bibinfo {title} {{Results from phase 1 of the HAYSTAC microwave
  cavity axion experiment}},\ }\href
  {https://doi.org/10.1103/PhysRevD.97.092001} {\bibfield  {journal} {\bibinfo
  {journal} {Phys. Rev. D}\ }\textbf {\bibinfo {volume} {97}},\ \bibinfo
  {pages} {092001} (\bibinfo {year} {2018})},\ \Eprint
  {https://arxiv.org/abs/1803.03690} {arXiv:1803.03690 [hep-ex]} \BibitemShut
  {NoStop}%
\bibitem [{\citenamefont {Backes}\ \emph {et~al.}(2021)\citenamefont {Backes}
  \emph {et~al.}}]{Backes:2020ajv}%
  \BibitemOpen
  \bibfield  {author} {\bibinfo {author} {\bibfnamefont {K.~M.}\ \bibnamefont
  {Backes}} \emph {et~al.} (\bibinfo {collaboration} {HAYSTAC}),\ }\bibfield
  {title} {\bibinfo {title} {{A quantum-enhanced search for dark matter
  axions}},\ }\href {https://doi.org/10.1038/s41586-021-03226-7} {\bibfield
  {journal} {\bibinfo  {journal} {Nature}\ }\textbf {\bibinfo {volume} {590}},\
  \bibinfo {pages} {238} (\bibinfo {year} {2021})},\ \Eprint
  {https://arxiv.org/abs/2008.01853} {arXiv:2008.01853 [quant-ph]} \BibitemShut
  {NoStop}%
\bibitem [{\citenamefont {Horns}\ \emph {et~al.}(2013)\citenamefont {Horns},
  \citenamefont {Jaeckel}, \citenamefont {Lindner}, \citenamefont {Lobanov},
  \citenamefont {Redondo},\ and\ \citenamefont {Ringwald}}]{Horns:2012jf}%
  \BibitemOpen
  \bibfield  {author} {\bibinfo {author} {\bibfnamefont {D.}~\bibnamefont
  {Horns}}, \bibinfo {author} {\bibfnamefont {J.}~\bibnamefont {Jaeckel}},
  \bibinfo {author} {\bibfnamefont {A.}~\bibnamefont {Lindner}}, \bibinfo
  {author} {\bibfnamefont {A.}~\bibnamefont {Lobanov}}, \bibinfo {author}
  {\bibfnamefont {J.}~\bibnamefont {Redondo}},\ and\ \bibinfo {author}
  {\bibfnamefont {A.}~\bibnamefont {Ringwald}},\ }\bibfield  {title} {\bibinfo
  {title} {{Searching for WISPy Cold Dark Matter with a Dish Antenna}},\ }\href
  {https://doi.org/10.1088/1475-7516/2013/04/016} {\bibfield  {journal}
  {\bibinfo  {journal} {JCAP}\ }\textbf {\bibinfo {volume} {1304}},\ \bibinfo
  {pages} {016}},\ \Eprint {https://arxiv.org/abs/1212.2970} {arXiv:1212.2970
  [hep-ph]} \BibitemShut {NoStop}%
\bibitem [{\citenamefont {Ahmed}\ \emph {et~al.}(2018)\citenamefont {Ahmed}
  \emph {et~al.}}]{Ahmed:2018oog}%
  \BibitemOpen
  \bibfield  {author} {\bibinfo {author} {\bibfnamefont {Z.}~\bibnamefont
  {Ahmed}} \emph {et~al.},\ }\bibfield  {title} {\bibinfo {title} {{Quantum
  Sensing for High Energy Physics}},\ }in\ \href
  {http://lss.fnal.gov/archive/2018/conf/fermilab-conf-18-092-ad-ae-di-ppd-t-td.pdf}
  {\emph {\bibinfo {booktitle} {{First workshop on Quantum Sensing for High
  Energy Physics}}}}\ (\bibinfo {year} {2018})\ \Eprint
  {https://arxiv.org/abs/1803.11306} {arXiv:1803.11306 [hep-ex]} \BibitemShut
  {NoStop}%
\bibitem [{\citenamefont {Kutas}\ \emph {et~al.}(2020)\citenamefont {Kutas},
  \citenamefont {Haase}, \citenamefont {Bickert}, \citenamefont {Riexinger},
  \citenamefont {Molter},\ and\ \citenamefont {von
  Freymann}}]{kutas2020terahertz}%
  \BibitemOpen
  \bibfield  {author} {\bibinfo {author} {\bibfnamefont {M.}~\bibnamefont
  {Kutas}}, \bibinfo {author} {\bibfnamefont {B.}~\bibnamefont {Haase}},
  \bibinfo {author} {\bibfnamefont {P.}~\bibnamefont {Bickert}}, \bibinfo
  {author} {\bibfnamefont {F.}~\bibnamefont {Riexinger}}, \bibinfo {author}
  {\bibfnamefont {D.}~\bibnamefont {Molter}},\ and\ \bibinfo {author}
  {\bibfnamefont {G.}~\bibnamefont {von Freymann}},\ }\bibfield  {title}
  {\bibinfo {title} {Terahertz quantum sensing},\ }\href
  {https://doi.org/10.1126/sciadv.aaz8065} {\bibfield  {journal} {\bibinfo
  {journal} {Sci. Adv.}\ }\textbf {\bibinfo {volume} {6}},\ \bibinfo {pages}
  {8065} (\bibinfo {year} {2020})}\BibitemShut {NoStop}%
\bibitem [{\citenamefont {{P. Barry et al.}}()}]{THzAxionSnowmassLoI}%
  \BibitemOpen
  \bibfield  {author} {\bibinfo {author} {\bibnamefont {{P. Barry et al.}}},\
  }\href@noop {} {\bibinfo {title} {{Opening the terahertz axion window}}},\
  \bibinfo {note}
  {\href{https://www.snowmass21.org/docs/files/summaries/CF/SNOWMASS21-CF2_CF0-AF7_AF0-IF1_IF2-UF2_UF0_Jesse_Liu-179.pdf}{Snowmass
  2021 Cosmic Frontier Letter of Interest No. 179}}\BibitemShut {NoStop}%
\bibitem [{\citenamefont {{A. Sonnenschein}}()}]{THzAxionSonnenschein}%
  \BibitemOpen
  \bibfield  {author} {\bibinfo {author} {\bibnamefont {{A. Sonnenschein}}},\
  }\href@noop {} {\bibinfo {title} {{Broadband Axion Searches with Coaxial Dish
  Antennas}}},\ \bibinfo {note}
  {\href{https://indico.fnal.gov/event/22434/contributions/205864/}{Axions
  beyond Gen 2 Workshop, Jan 2021}}\BibitemShut {NoStop}%
\bibitem [{\citenamefont {{Thorlabs Inc.}}()}]{ThorlabsKinesis}%
  \BibitemOpen
  \bibfield  {author} {\bibinfo {author} {\bibnamefont {{Thorlabs Inc.}}},\
  }\href@noop {} {\bibinfo {title} {{Kinesis software}}},\ \bibinfo {note}
  {\href{https://www.thorlabs.com/software_pages/ViewSoftwarePage.cfm?Code=Motion_Control&viewtab=0}{KDC101
  K-Cube Motor Controller}}\BibitemShut {NoStop}%
\bibitem [{\citenamefont {{Gentec Electro-Optics}}()}]{Gentec-THZ-BL-curves}%
  \BibitemOpen
  \bibfield  {author} {\bibinfo {author} {\bibnamefont {{Gentec
  Electro-Optics}}},\ }\href@noop {} {}\bibinfo {note} {\href{
  https://downloads.gentec-eo.com/prod/104e85b0/Curves_Terahertz_2020_V1.0_EN.pdf
  }{THZ-BL Absorption Curves}}\BibitemShut {NoStop}%
\bibitem [{\citenamefont {Harris}\ \emph {et~al.}(2020)\citenamefont {Harris}
  \emph {et~al.}}]{2020NumPy-Array}%
  \BibitemOpen
  \bibfield  {author} {\bibinfo {author} {\bibfnamefont {C.~R.}\ \bibnamefont
  {Harris}} \emph {et~al.},\ }\bibfield  {title} {\bibinfo {title} {Array
  programming with {NumPy}},\ }\href
  {https://doi.org/10.1038/s41586-020-2649-2} {\bibfield  {journal} {\bibinfo
  {journal} {Nature}\ }\textbf {\bibinfo {volume} {585}},\ \bibinfo {pages}
  {357–362} (\bibinfo {year} {2020})}\BibitemShut {NoStop}%
\bibitem [{\citenamefont {{W}es
  {M}c{K}inney}(2010)}]{mckinney-proc-scipy-2010}%
  \BibitemOpen
  \bibfield  {author} {\bibinfo {author} {\bibnamefont {{W}es {M}c{K}inney}},\
  }\bibfield  {title} {\bibinfo {title} {{D}ata {S}tructures for {S}tatistical
  {C}omputing in {P}ython},\ }in\ \href
  {https://doi.org/10.25080/Majora-92bf1922-00a} {\emph {\bibinfo {booktitle}
  {{P}roceedings of the 9th {P}ython in {S}cience {C}onference}}},\ \bibinfo
  {editor} {edited by\ \bibinfo {editor} {\bibnamefont {{S}t\'efan van~der
  {W}alt}}\ and\ \bibinfo {editor} {\bibnamefont {{J}arrod {M}illman}}}\
  (\bibinfo {year} {2010})\ pp.\ \bibinfo {pages} {56 -- 61}\BibitemShut
  {NoStop}%
\bibitem [{\citenamefont {{Hunter}}(2007)}]{matplotlib}%
  \BibitemOpen
  \bibfield  {author} {\bibinfo {author} {\bibfnamefont {J.~D.}\ \bibnamefont
  {{Hunter}}},\ }\bibfield  {title} {\bibinfo {title} {Matplotlib: A 2d
  graphics environment},\ }\href {https://doi.org/10.1109/MCSE.2007.55}
  {\bibfield  {journal} {\bibinfo  {journal} {Computing in Science
  Engineering}\ }\textbf {\bibinfo {volume} {9}},\ \bibinfo {pages} {90}
  (\bibinfo {year} {2007})}\BibitemShut {NoStop}%
\bibitem [{\citenamefont {Virtanen}\ \emph {et~al.}(2020)\citenamefont
  {Virtanen} \emph {et~al.}}]{2020SciPy-NMeth}%
  \BibitemOpen
  \bibfield  {author} {\bibinfo {author} {\bibfnamefont {P.}~\bibnamefont
  {Virtanen}} \emph {et~al.},\ }\bibfield  {title} {\bibinfo {title} {{{SciPy}
  1.0: Fundamental Algorithms for Scientific Computing in Python}},\ }\href
  {https://doi.org/10.1038/s41592-019-0686-2} {\bibfield  {journal} {\bibinfo
  {journal} {Nat. Methods}\ }\textbf {\bibinfo {volume} {17}},\ \bibinfo
  {pages} {261} (\bibinfo {year} {2020})}\BibitemShut {NoStop}%
\bibitem [{\citenamefont {Welch}(1967)}]{welch1967use}%
  \BibitemOpen
  \bibfield  {author} {\bibinfo {author} {\bibfnamefont {P.}~\bibnamefont
  {Welch}},\ }\bibfield  {title} {\bibinfo {title} {The use of fast fourier
  transform for the estimation of power spectra: a method based on time
  averaging over short, modified periodograms},\ }\href@noop {} {\bibfield
  {journal} {\bibinfo  {journal} {IEEE Transactions on audio and
  electroacoustics}\ }\textbf {\bibinfo {volume} {15}},\ \bibinfo {pages} {70}
  (\bibinfo {year} {1967})}\BibitemShut {NoStop}%
\bibitem [{\citenamefont {{HawkEye Technologies, LLC and Boston
  Electronics}}()}]{BE-IR-Si253}%
  \BibitemOpen
  \bibfield  {author} {\bibinfo {author} {\bibnamefont {{HawkEye Technologies,
  LLC and Boston Electronics}}},\ }\href@noop {} {}\bibinfo {note} {\href{
  https://www.boselec.com/wp-content/uploads/Linear/IRSources/IRSourcesLiterature/IR-Si253-Series-formerly-IR-18-and-IR-19.pdf
  }{IR-Si253 Engineering Data Charts}}\BibitemShut {NoStop}%
\bibitem [{\citenamefont {Rajandas}\ \emph {et~al.}(2012)\citenamefont
  {Rajandas}, \citenamefont {Parimannan}, \citenamefont {Sathasivam},
  \citenamefont {Ravichandran},\ and\ \citenamefont {{Su Yin}}}]{PEW_plot}%
  \BibitemOpen
  \bibfield  {author} {\bibinfo {author} {\bibfnamefont {H.}~\bibnamefont
  {Rajandas}}, \bibinfo {author} {\bibfnamefont {S.}~\bibnamefont
  {Parimannan}}, \bibinfo {author} {\bibfnamefont {K.}~\bibnamefont
  {Sathasivam}}, \bibinfo {author} {\bibfnamefont {M.}~\bibnamefont
  {Ravichandran}},\ and\ \bibinfo {author} {\bibfnamefont {L.}~\bibnamefont
  {{Su Yin}}},\ }\bibfield  {title} {\bibinfo {title} {A novel ftir-atr
  spectroscopy based technique for the estimation of low-density polyethylene
  biodegradation},\ }\href
  {https://doi.org/https://doi.org/10.1016/j.polymertesting.2012.07.015}
  {\bibfield  {journal} {\bibinfo  {journal} {Polym. Test.}\ }\textbf {\bibinfo
  {volume} {31}},\ \bibinfo {pages} {1094} (\bibinfo {year}
  {2012})}\BibitemShut {NoStop}%
\end{thebibliography}%

\appendix 

\section{\label{sec:Fourier}FTS Spectrum Extraction}

\begin{figure}
\centering
\includegraphics[width=\linewidth]{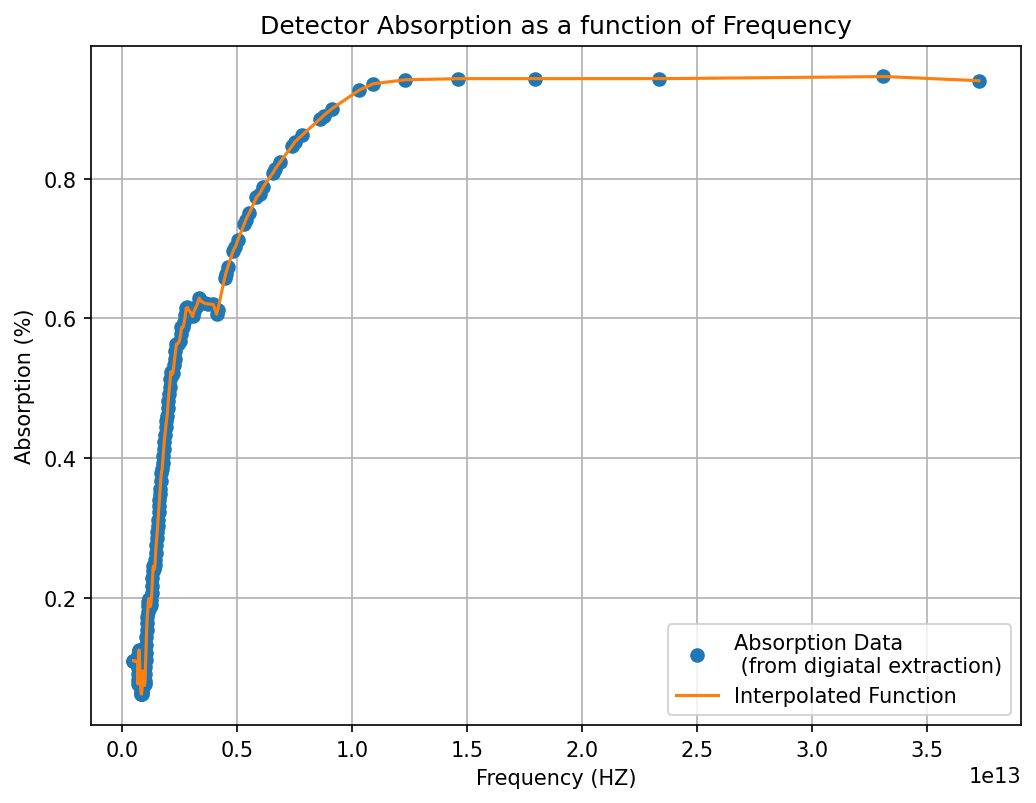}
\caption{\label{fig:GentecAbsorption}Frequency dependence of absorption response for the Gentec photodetector using the manufacturer datasheet~\cite{Gentec-THZ-BL-curves}.  }
\end{figure}

This appendix briefly reviews Fourier transform methods together with the implemented simulations and power spectra measurements. 
The power of an ideal monochromatic source with wavenumber $k = 1/\lambda = f/c$ is recorded by the detector as a function of the mirror displacement $x$ gives the interferogram
\begin{equation}
    p(x) = \frac{p_0}{2}\left[1+\cos(2\pi k x)\right],\label{eq:even_interferogram}
\end{equation}
where $p_0$ is the constant unmodulated power.
For a general power spectrum $p(k)$, the interferogram is the integral over all spectral wavenumbers
\begin{equation}
    p(x) = \frac{1}{2}\int_0^{\infty}p(k)\left[1+\cos(2\pi k x)\right]dk \label{eq:exactInt}.
\end{equation}
This can be discretized using finite wavenumber segments $\Delta k$ in the following way:
\begin{align}
    p(x) &\approx \sum_{i=0}^\infty \left[\int_{i\Delta k}^{(i+1)\Delta k}\!\!\! p(k)dk\right] \frac{1+\cos\left(2\pi \Delta k  (i+\frac{1}{2}) x\right)}{2} \nonumber\\
    &= \sum_{i=0}^\infty  w_i\frac{1+\cos\left(2\pi  k_ix\right)}{2},
\end{align}
where $k_i = (i+\frac{1}{2})\Delta k$ and $w_i$ are weights defined by
\begin{equation}
    w_i = \int_{i\Delta k}^{(i+1)\Delta k}p(k)dk \approx p(k_i)\Delta k.
\end{equation}
The last approximation is true for linear functions or small integration ranges but is a biased, if simplified, approximation for exponential functions. 
For a sufficiently small integration step size, this approximate form is appropriate and gives  the sum
\begin{equation}
    p(x)\approx \sum_{i=0}^\infty   \frac{p(k_i)}{2}\left[1+\cos\left(2\pi k_i x \right)\right] \Delta k.
    \label{input_signal_FTS}
\end{equation}
which reduces to the integral from Eq.~\eqref{eq:exactInt} in the limit $\Delta k\rightarrow 0$. 
This discretization prescription is used to implement the simulation of input power, for example taking $p(k)$ as predicted by Planck's law for broadband radiation. 
We can account for effects such as digitization noise due to the analogue-to-digital converters in the simulations. 
For simplicity, many expected instrument and environmental effects are not included for our simple simulations, such as beam divergence and atmospheric absorption.
Furthermore, Figure~\ref{fig:GentecAbsorption} shows the manufacturer-supplied response absorption efficiency for the Gentec detector as a function of frequency. 
For the frequencies considered by this paper $f \gtrsim 10$~THz, the absorption efficiency is relatively constant above 90\%. 
For $f \lesssim 10$~THz, this absorption factor drops quickly falling to around 60\% at 3~THz and 10~\% around 1~THz. 
For shape comparisons between the measurements and the spectra predicted by Planck's law in the main text, this suppressed absorption at low frequencies is not considered in our simulation.

\begin{figure}
\centering
\includegraphics[width=\linewidth]{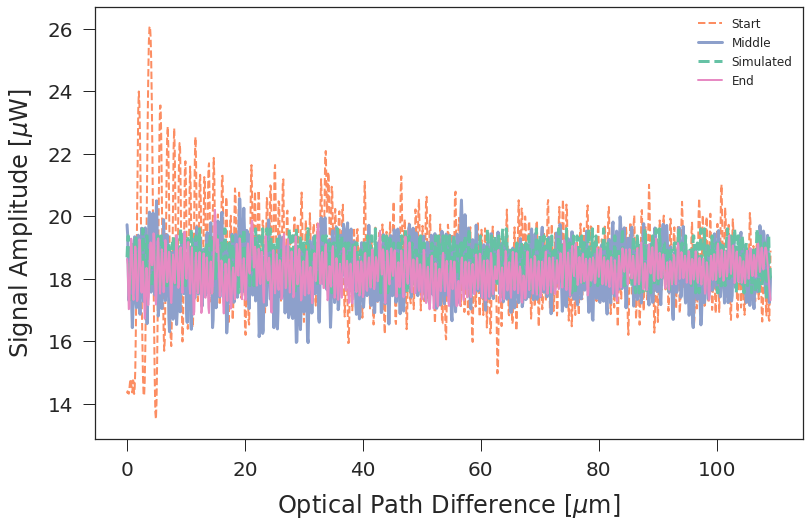}\\
\includegraphics[width=\linewidth]{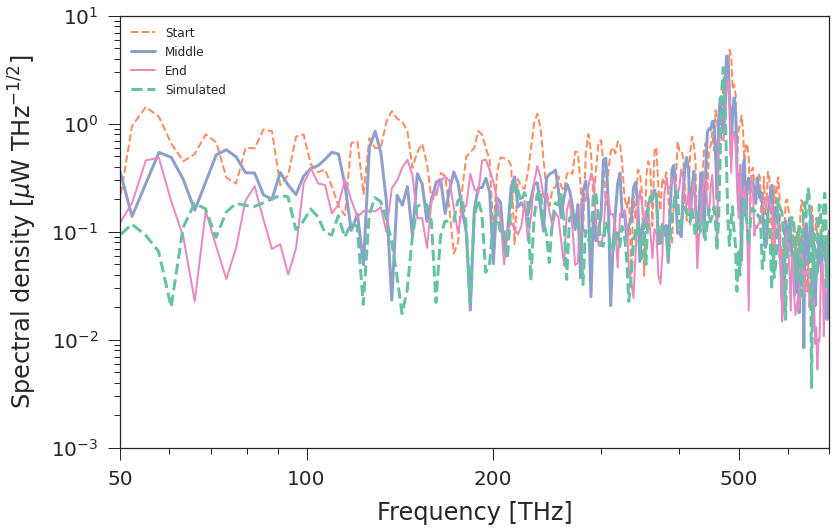}
\caption{\label{fig:laser_3periods} Initial  stability check of the FTS after initial alignment using 635 nm laser: (upper) interferogram power vs displacement and (lower) frequency space. }
\end{figure}

\begin{figure}
\centering
\includegraphics[width=\linewidth]{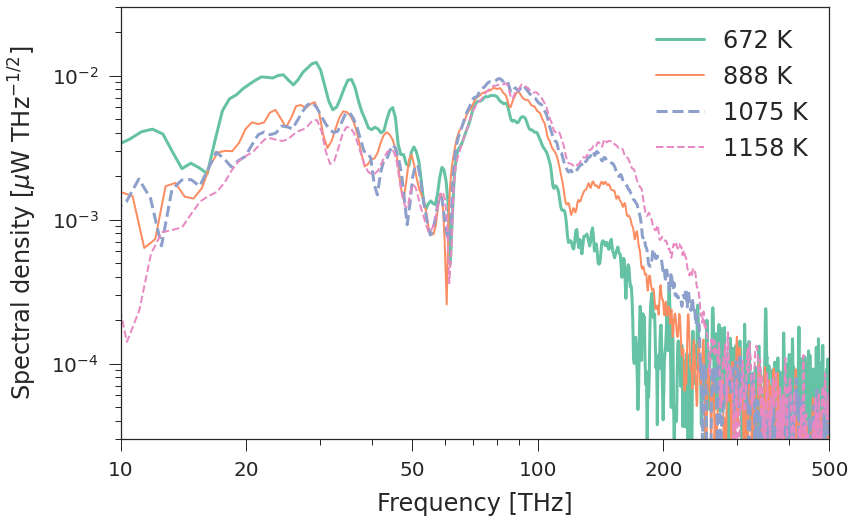}
\caption{\label{fig:freq_vs_temp}Normalized power spectral density of IR-Si253 source with varied voltage settings corresponding to different temperatures. The spectra are normalized such that the area under each curve is the same to facilitate shape comparison.}
\end{figure}

Turning to the power spectrum in wavenumber space $p(k)$, this is formally the Fourier transform of the input interferogram distribution
\begin{equation}
    p(k) = \mathcal{F}[p(x)] = \int_{-\infty}^\infty p(x)e^{2 \pi i k x }. 
\end{equation}
Practically, this is implemented numerically by considering $N$ segments of the interferogram $p(x)$ labelled by $n$ and computing the corresponding periodogram value $P_n(k)$. 
This is defined as the discrete Fourier transform weighted by a windowing function $W(x_m)$ which we select to be \texttt{parzen} in the \textsc{SciPy} implementation
\begin{equation}
    P_n(k) = \frac{\left|\sum_{m=1}^M p(x_m)W(x_m) e^{2 \pi i k m / M }
\right|^2}{\sum_{m=1}W^2(x_m)}.
\end{equation}
The estimated power spectral density $p_\text{psd}(k)$ is then the average over the $N$ segments of the periodograms
\begin{equation}
    p_\text{psd}(k) = \left|\frac{1}{N}\sum_{n=1}^N P_n(k)\right|^{1/2}.
\end{equation}
Further details of the implementation can be found in the \texttt{signal.periodogram} documentation of the \textsc{SciPy} package~\cite{2020SciPy-NMeth}.

\section{Additional studies}

Figure~\ref{fig:laser_3periods} shows the interferogram  using the 635 nm laser partitioning the data into three parts.
This was performed as to verify the stability of the motorized stage and detector measurements after the initial laser-based alignment in preparation for nominal operation using our interferometer.

Figure~\ref{fig:freq_vs_temp} shows the normalized power spectra of the IR-Si253 source for different voltage settings. 
The spectra are normalized to the same area under each curve to compare the shapes between the temperature. 
We observe that increasing temperature indeed increases the power at the high frequencies, as one would expect from Planck's law. 
In particular, the exponential suppression at high frequencies moves from around 120~THz to 200~THz from 672~K to 1158~K, respectively. 
This verifies the broadband spectral capabilities of our FTS, and allows us to select the lowest temperature that would enable use of the 1650~nm bandpass filter at 182~THz. 

\begin{figure}[tbh]
\centering
\includegraphics[width=\linewidth]{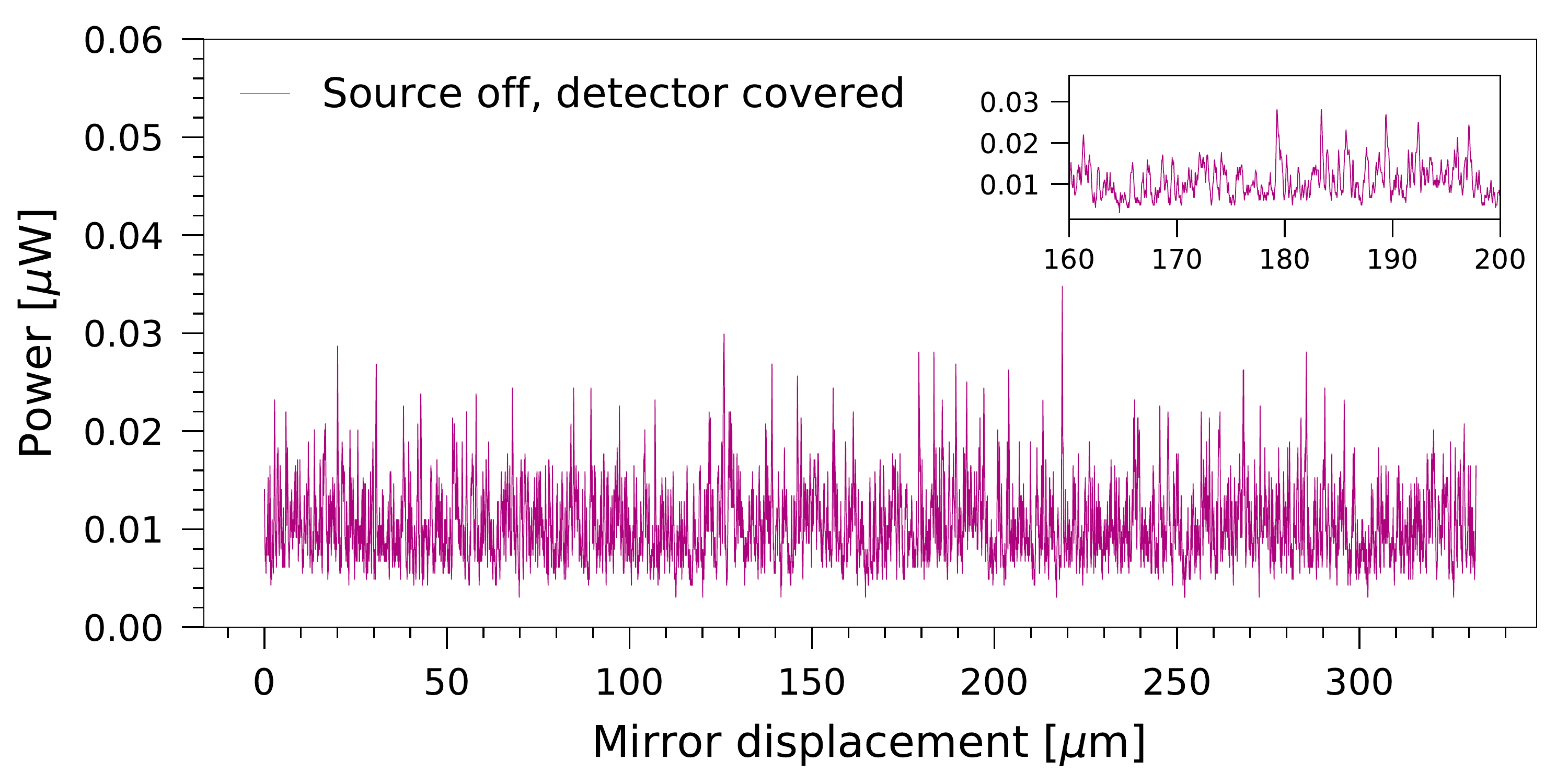}
\includegraphics[width=\linewidth]{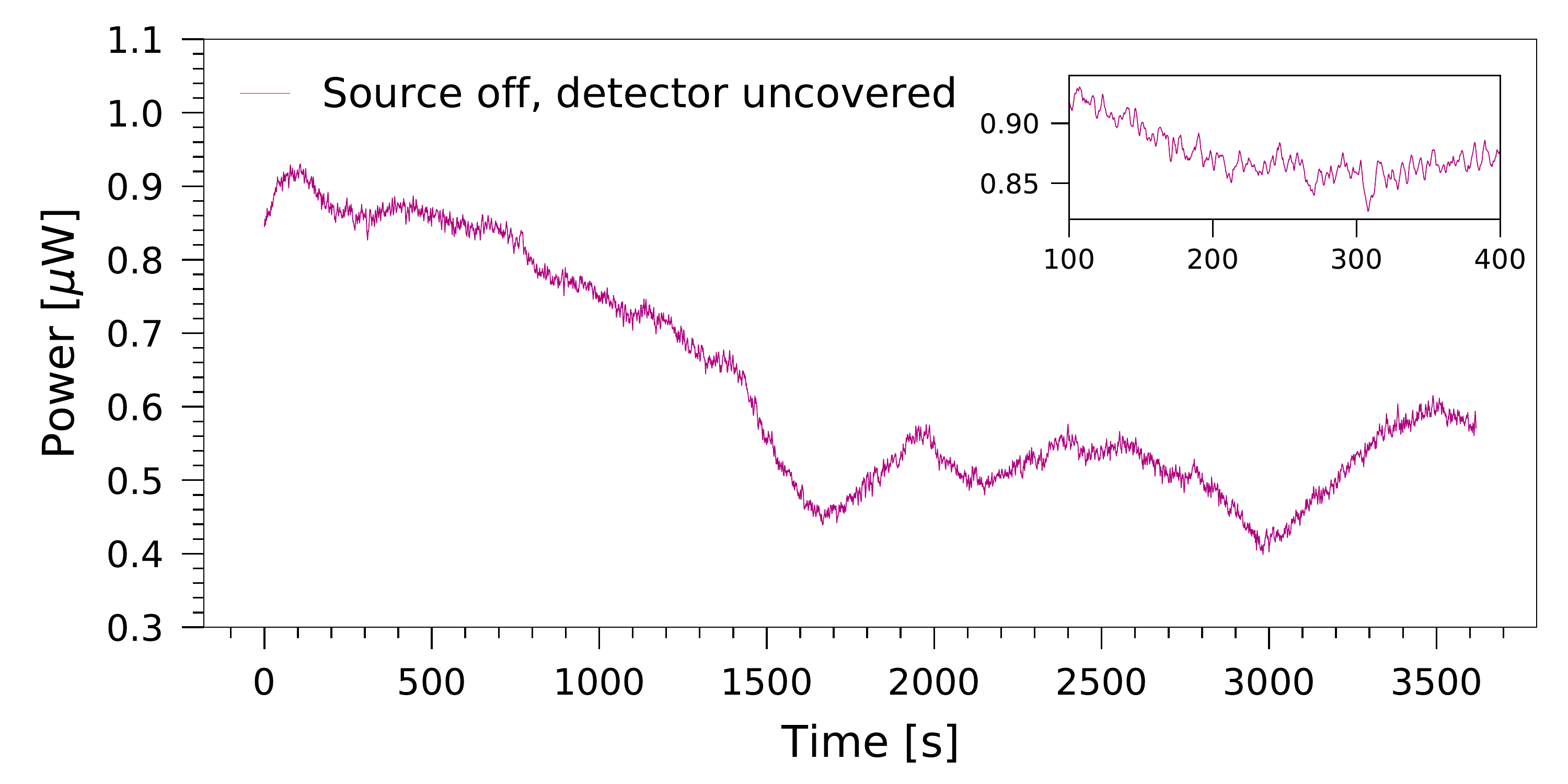}
\caption{\label{fig:AmbientNoise}Noise measurements with source switched off for (top) detector covered, moving mirror and (bottom) detector uncovered, stationary mirror. The inset plot zooms into a small subset of the displayed range. These were taken with the maximum power on the Gentec detector set to 2~$\mu$W.}
\end{figure}

Figure~\ref{fig:AmbientNoise} displays ambient noise measurements from the photosensor with the sources switched off.
First in Fig.~\ref{fig:AmbientNoise} (top), the detector was obscured with the cover supplied with the detector by Gentec while moving the mirror at 100~nm~s$^{-1}$ as done in nominal data taking. 
The measurements are taken with the lowest saturation power scale set to $2~\mu$W, which minimizes the ADC digitization noise floor. 
We observe a minimum noise power range [5, 30]~nW compared with the manufacturer specification of 50~nW, providing an indication of the intrinsic detector-readout noise. 
Then in Fig.~\ref{fig:AmbientNoise} (bottom), the manufacturer-supplied cover was removed with the chopper switched on to measure noise power contributions from the environment for one hour.
We find that the noise is at least an order of magnitude higher, and has high-frequency stochastic variations (inset plot) in addition to low-frequency drifts on the order of hundreds of seconds.
This provides an initial characterization of the baseline environmental noise when operating the nominal data taking with the source swtiched on.

\end{document}